\documentclass[twocolumn,twocolappendix]{aastex701}
\usepackage{xspace}
\usepackage{amsmath}

\shorttitle{Where do stars explode?}
\shortauthors{Sarbadhicary et al.}

\begin{document}
\newcommand{\vdag}{(v)^\dagger}
\newcommand\aastex{AAS\TeX}
\newcommand\latex{La\TeX}
\newcommand{\Msun}{M$_{\odot}$}	
\newcommand{\Mdot}{M$_{\odot}$ yr$^{-1}$\xspace}
\newcommand{\kms}{km s$^{-1}$\xspace}
\newcommand{\Msunpc}{M$_{\odot}$ pc$^{-2}$\xspace}
\newcommand{\fmol}{$f_{\mathrm{H_2}}$\xspace}
\newcommand{\hi}{\ion{H}{1}\xspace}
\newcommand{\hii}{\ion{H}{2}\xspace}
\newcommand{\cmc}{cm$^{-3}$\xspace}
\newcommand{\mum}{$\mu$m\xspace}
\newcommand{\cmipsunit}{(erg s$^{-1}$)(M$_{\odot}$ yr$^{-1}$)$^{-1}$\xspace}
\newcommand{\revise}[1]{\textbf{#1}\xspace}

\title{Where do stars explode in the ISM? -- The distribution of dense gas around evolved massive stars in M33}

\author[0000-0002-4781-7291]{Sumit K. Sarbadhicary}
\affiliation{Department of Physics, The Ohio State University, Columbus, Ohio 43210, USA}
\affiliation{Center for Cosmology \& Astro-Particle Physics, The Ohio State University, Columbus, Ohio 43210, USA}
\affiliation{Department of Astronomy, The Ohio State University, Columbus, Ohio 43210, USA}
\email[show]{ssarbad1@jh.edu}

\author{Jordan Wagner}
\affiliation{Department of Physics, The Ohio State University, Columbus, Ohio 43210, USA}
\affiliation{Center for Cosmology \& Astro-Particle Physics, The Ohio State University, Columbus, Ohio 43210, USA}
\affiliation{Department of Astronomy, The Ohio State University, Columbus, Ohio 43210, USA}
\email{wagner.1453@buckeyemail.osu.edu}

\author[0000-0001-9605-780X]{Eric W. Koch}
\affiliation{Center for Astrophysics $\mid$ Harvard \& Smithsonian, 60 Garden St., 02138 Cambridge, MA, USA}
\email{koch.eric.w@gmail.com}

\author[0000-0002-5993-6685]{Ness Mayker Chen}
\affiliation{Center for Cosmology \& Astro-Particle Physics, The Ohio State University, Columbus, Ohio 43210, USA}
\affiliation{Department of Astronomy, The Ohio State University, Columbus, Ohio 43210, USA}
\email{maykerchen.1@osu.edu}

\author[0000-0002-2545-1700]{Adam K. Leroy}
\affiliation{Center for Cosmology \& Astro-Particle Physics, The Ohio State University, Columbus, Ohio 43210, USA}
\affiliation{Department of Astronomy, The Ohio State University, Columbus, Ohio 43210, USA}
\email{leroy.42@osu.edu}

\author[0000-0003-2166-1935]{Natalia Lah\'{e}n}
\affiliation{Max-Planck-Institute f\"{u}r Astrophysik, Karl-Schwarzschild-Stra{\ss}e 1, D-85740 Garching, Germany}
\email{nlahen@mpa-garching.mpg.de}

\author[0000-0002-5204-2259]{Erik~Rosolowsky}
\affiliation{Department of Physics, University of Alberta, Edmonton, AB T6G 2E1, Canada}
\email{rosolowsky@ualberta.ca}

\author[0000-0002-5787-138X]{Kathryn F. Neugent}
\altaffiliation{NASA Hubble Fellow}
\affiliation{Center for Astrophysics $\mid$ Harvard \& Smithsonian, 60 Garden St., 02138 Cambridge, MA, USA}
\email{kathryn.neugent@cfa.harvard.edu}

\author[0000-0003-2896-3725]{Chang-Goo Kim}
\affiliation{Department of Astrophysical Sciences, Princeton University, Princeton, NJ 08544, USA}
\email{cgkim@astro.princeton.edu}

\author[0000-0002-8400-3705]{Laura Chomiuk}
\affiliation{Center for Data Intensive and Time Domain Astronomy, Department of Physics and Astronomy, Michigan State University, East Lansing, MI 48824, USA}
\email{chomiuk@pa.msu.edu}

\author[0000-0002-1264-2006]{Julianne J. Dalcanton}
\affiliation{Department of Astronomy, University of Washington, Box 351580, Seattle, WA 98195-1580}
\affiliation{Center for Computational Astrophysics, Flatiron Institute, 162 Fifth Ave, New York, NY 10010, USA}
\email{jdalcanton@flatironinstitute.org}

\author[0000-0002-1790-3148]{Laura A. Lopez}
\affiliation{Center for Cosmology \& Astro-Particle Physics, The Ohio State University, Columbus, Ohio 43210, USA}
\affiliation{Department of Astronomy, The Ohio State University, Columbus, Ohio 43210, USA}
\email{lopez.513@osu.edu}

\author[0000-0001-9504-7386]{Nickolas M. Pingel}
\affiliation{Department of Astronomy, University of Wisconsin-Madison, 475 North Charter St., Madison, WI, 53706-15821, USA}
\email{nmpingel@wisc.edu}

\author[0000-0002-4663-6827]{Remy Indebetouw}
\affiliation{University of Virginia Astronomy, Dept. 530 McCormick Road, 
Charlottesville, VA 22904-4325}
\affiliation{National Radio Astronomy Observatory,  520 Edgemont Road, 
Charlottesville, VA 22903-2475}
\email{remy@virginia.edu}

\author[0000-0002-0012-2142]{Thomas G. Williams}
\email{thomas.williams@physics.ox.ac.uk}
\affiliation{Sub-department of Astrophysics, Department of Physics, University of Oxford, Keble Road, Oxford OX1 3RH, UK}

\author[0000-0003-1356-1096]{Elizabeth Tarantino}
\affiliation{Space Telescope Science Institute, 3700 San Martin Drive, Baltimore, MD 21218, USA}
\affiliation{Department of Astronomy, University of Maryland, College Park, MD 20742, USA}
\email{etarantino@stsci.edu}

\author[0000-0002-3106-7676]{Jennifer Donovan Meyer}
\email{jmeyer@nrao.edu}
\affiliation{National Radio Astronomy Observatory,  520 Edgemont Road, 
Charlottesville, VA 22903-2475}

\author[0000-0003-0605-8732]{Evan D. Skillman}
\email{skillman@astro.umn.edu}
\affiliation{Minnesota Institute for Astrophysics, University of Minnesota, Minneapolis, MN 55455, USA}

\author[0000-0003-2599-7524]{Adam Smercina}
\affiliation{Department of Astronomy, University of Washington, Box 351580, Seattle, WA 98195-1580}
\email{asmerci@uw.edu*}

\author[0000-0002-3227-4917]{Amanda A.\,Kepley}
\email{akepley@nrao.edu}
\affiliation{National Radio Astronomy Observatory,  520 Edgemont Road, Charlottesville, VA 22903-2475}

\author[0000-0001-7089-7325]{Eric J.\,Murphy}
\email{emurphy@nrao.edu}
\affiliation{National Radio Astronomy Observatory,  520 Edgemont Road, 
Charlottesville, VA 22903-2475}

\author[0000-0002-1468-9668]{Jay Strader}
\affiliation{Center for Data Intensive and Time Domain Astronomy, Department of Physics and Astronomy, Michigan State University, East Lansing, MI 48824, USA}
\email{straderj@msu.edu}

\author[0000-0002-7759-0585]{Tony Wong}
\affiliation{Department of Astronomy, University of Illinois, Urbana, IL 61801, USA}
\email{wongt@illinois.edu}

\author[0000-0002-3418-7817]{Sne{\v{z}}ana Stanimirovi{\'c}}
\affiliation{Department of Astronomy, University of Wisconsin-Madison, 475 North Charter St., Madison, WI, 53706-15821, USA}
\email{sstanimi@astro.wisc.edu}

\author[0000-0002-5877-379X]{Vicente Villanueva}
\affiliation{Department of Astronomy, University of Maryland, College Park, MD 20742, USA}
\email{vicente.villanueva@postgrado.uv.cl}

\author[0000-0003-4793-7880]{Fabian Walter}
\affiliation{Max Planck Institute for Astronomy, K\"{o}nigstuhl 17, D-69117 Heidelberg, Germany}
\affiliation{National Radio Astronomy Observatory, Pete V. Domenici Array Science Center, P.O. Box O, Socorro, NM 87801, USA}
\email{fabian.walter@aasjournals.org}

\author[0000-0001-8224-1956]{Juergen Ott}
\affiliation{New Mexico Institute of Mining and Technology, 801 Leroy Place, Socorro, NM 87801, USA}
\affiliation{National Radio Astronomy Observatory, Pete V. Domenici Array Science Center, P.O. Box O, Socorro, NM 87801, USA}
\email{jott@nrao.edu}

\author[0000-0003-2511-2060]{Jeremy Darling}
 \affiliation{Center for Astrophysics and Space Astronomy,
Department of Astrophysical and Planetary Sciences,
University of Colorado, 389 UCB,
Boulder, CO 80309-0389, USA}
\email{jeremy.darling@colorado.edu}

\author[0000-0001-6326-7069]{Julia Roman-Duval}
\affiliation{Space Telescope Science Institute,
3700 San Martin Drive,
Baltimore, MD 21218, USA}
\email{duval@stsci.edu}

\author[0000-0002-7743-8129]{Claire E. Murray}
\affiliation{Space Telescope Science Institute,
3700 San Martin Drive,
Baltimore, MD 21218, USA}
\affil{Department of Physics \& Astronomy, 
Johns Hopkins University,
3400 N. Charles Street, 
Baltimore, MD 21218}
\email{cmurray1@stsci.edu}

\begin{abstract}
The effect of supernovae (SNe) on star-formation in the interstellar medium (ISM) depends sensitively on where SNe explode with respect to ISM clouds. Observationally, SN ISM environments characterized by spatially-resolved gas maps can empirically guide the placement of SNe in subgrid models, but unfortunately such measurements remain scarce, as SNe are rare and often distant. Here we demonstrate a new approach -- mapping the ISM around evolved massive stars that are soon to explode. These provide a substantially larger sample of `explosion sites' (than just historical SNe) in nearby galaxies that have high-resolution atomic and molecular ISM maps from Jansky VLA and ALMA. We demonstrate this technique in the well-resolved Local Group spiral M33 by analyzing the 50 pc-scale projected ISM densities around red supergiants (RSGs, 8-30 \Msun stars) Wolf-Rayet stars (WRs, $>$30\Msun stars), and supernova remnants (SNRs). We find a \emph{mass-dependent} correlation between stars and gas clouds, with atleast 45\% of WRs and upto 77\% of RSGs having no detectable H$_2$ at their pixel locations. In the sample with H$_2$ detections, we find that more massive younger progenitors are coincident with denser gas. We show that the density distributions for stars $>$15 \Msun are statistically distinct from random alignment of stars and gas in M33. Our work provides the first \emph{observationally-derived} estimate of the fraction of the SN-producing stellar population correlated with ISM density peaks. We demonstrate how this can be compared with galaxy simulations, and advocate similar comparisons to the community for constraining sub-grid models.
\end{abstract}


\section{Introduction} \label{sec:intro}
Feedback from supernovae (SNe) plays a central role in galaxy evolution. SN explosions drive outflows \citep[e.g.][]{Strickland2009, Heckman2017, Veilleux2020}, turbulence and the multi-phase distribution of the interstellar medium \citep[ISM, e.g.][]{MO77, Joung2006, Ostriker2011, Hill2012, Kim2013, Martizzi2016, Kannan2020, Rathjen2021, Kim2022}, accelerate cosmic rays \citep[e.g.][]{Ackermann2013, Caprioli2014, Bykov2018}, and disperse metals synthesized in stars \citep[see e.g.][for recent simulations and measurements]{Walch2015, Andrews2017, Telford2019, Li2020}. The balance of these mechanisms determines the star-forming properties of the ISM \citep[as shown e.g. in the recent simulations of][one of many examples]{Ostriker2022}.

Despite its importance, many uncertainties remain in the physics of SN feedback in simulations, a major one being their locations with respect to ISM density peaks. Densities and temperatures in the ISM spans several orders of magnitudes, and these affect the spatial scales and timescales on which SN blastwaves cool and subsequently share the thermal energy and momentum with the ambient gas \citep{Li2015, Kim2015, Martizzi2015, Karpov2020, Koo2020, Osaka}. As a result, the accumulation of thermal energy and momentum from SNe can significantly vary with the environments in which they are seeded. Simulations have confirmed the importance of the SN ambient density, finding large differences in the global properties of the ISM such as star-formation, outflow rates and phase distribution, depending on the correlation between SN locations and ISM density peaks \citep[e.g]{Gatto2015,Walch2015,Girichidis2016,Martizzi2016}.

Unfortunately, the locations where stars explode depend on a number of poorly understood physical properties of massive stars, such as their evolutionary timescales \citep[e.g.,][]{Zapartas2017, Eldridge2022}, potential for explosion from different progenitor masses $\gtrsim$18 \Msun \citep{Sukhbold2016}, single/binary-evolution-driven mass-loss \citep{Langer2012, Smith2014}, 3D kinematics from interactions in many-body environments \citep{Oh2016, Oey2018} or kicks from SN explosions in binary systems \citep{Eldridge2011, Renzo2019, Dallas2022}, and the co-evolution timescales with their parent molecular clouds \citep[e.g.,][]{Kruijssen2019, Chevance2020}. While modern high-resolution simulations are increasingly making great strides in capturing many of these intricate processes and their effects on the ISM \citep[e.g.][]{Kim2017b, Hu2017, Gutcke2021}, a comprehensive set of observations of the diverse ambient environments of SNe is needed to serve as a useful guide for simulations, but presently lacking.

 In recent years, detailed mapping of the multi-phase ISM down to scales ($\sim$100 pc) of individual clouds and their co-eval stellar associations have become possible with modern synthesis arrays like the Atacama Large Millimeter Array (ALMA) and the Karl G.\ Jansky Very Large Array (VLA) -- effectively opening the window to \emph{direct} observations of the cloud-scale ISM around SNe. \cite{MC22} (hereafter MC22) studied the molecular gas environments of 63 historical SNe in 31 galaxies at physical resolutions spanning 60-150 pc, using the PHANGS survey of nearby galaxies within 20 Mpc \citep{Leroy2021}. They found about 15$\%$ of stripped envelope (SE) SNe (Type IIb, Ib, Ic) and 50$\%$ of Type II SNe explode at least 150 pc away from the sites of giant molecular clouds (GMCs). This was a \emph{direct} confirmation that a significant fraction (if not all) core-collapse SNe explode in low-density environments outside of GMCs. The work also greatly benefited from the cloud-scale resolution that was inaccessible to previous generation extragalactic CO-surveys \citep{Galbany2017}.

The potential to expand such SN location analyses however, faces a more fundamental limitation --- SNe in individual galaxies are rare and require observing large cosmological volumes to obtain a sizable sample \citep{Li2011}, precluding the high spatial resolution needed to characterize their environments. While the mm-wavelengths are short enough that ALMA can still resolve molecular clouds out to 20 Mpc with longer baselines, the ability to observe the atomic (\hi) ISM -- which is a significant component by both mass and volume in star-forming galaxies \citep{Kulkarni1987, Dickey1990, Walter2008} -- at arcsecond resolution out to the same distances is unfeasible, at least until the advent of the next-generation VLA \citep{Murphy2018} and Square Kilometer Array (SKA). Thus while studies of the molecular gas around SNe are promising in the nearby universe, the \emph{total} ISM density at the scale of individual clouds around SNe restricts our observing volume to the nearest galaxies, where again, SN counts become vanishingly small.

Here we propose a novel strategy for uncovering the ISM distribution near SNe -- by focusing instead on the Local Group galaxies, and their substantial population of evolved massive stars that will likely explode within a few Myr. The approach has two major advantages. Firstly, even though SNe are rare, we have access to high-completeness catalogs of thousands of evolved massive stars like red supergiants (RSGs) and Wolf-Rayet stars \citep[WRs,][]{Neugent2011,Massey2021} that have been meticulously constructed over decades and are strong candidates for future core-collapse SNe (see Section \ref{sec:obs-why}). We also have access to supernova remnants \citep[SNRs,][]{White2019} which trace locations of actual explosions that have occurred in the past. The combination of WRs, RSGs and SNRs enables a much larger catalog of `explosion sites' than achievable with a SN catalog while retaining substantial spatial detail of the environments. Secondly, the proximity of Local Group galaxies\footnote{While not the focus of this paper, we note that another advantage in the Local Group is availability of detailed \emph{star-formation histories} from observations of resolved stellar populations \citep{Massey2006, Cioni2011, Dalcanton2012, Nidever2017, Lazzarini2022}. We use them briefly in our analyses to obtain useful insight into the distribution of our massive stars in certain sections.} offers not only cloud-scale observations of molecular gas \citep{Engargiola2003, Rosolowsky2007, Miura2012,Wong2011,Druard2014,Caldu2015}, but \emph{also} the atomic gas phase \citep[e.g.][]{Braun2009, Braun2012, Koch2018, Pingel2022}, with detailed information about kinematics, opacity and the cold/warm phases that is not as reliably recovered in more distant targets. 

We carry out our work with published and new datasets of M33, and organize our paper as follows. In Section \ref{sec:obs}, we describe the published catalogs of RSGs, WRs and SNRs that we use, present our new observations of \hi and CO with the VLA and ALMA, and elaborate on our overall strategy of observing the ISM around evolved massive stars. Section \ref{sec:results} describes our results, including the relative spatial distribution of the stars and gas, and the density distributions; we statistically verify these densities with multi-wavelength star-formation maps. Section \ref{sec:discussion} discusses the implications of the results, specifically the correlation between stellar age and ambient ISM density, the significant fraction of massive stars in the low-density atomic ISM, the need for higher ($\sim$pc) resolution molecular data to identify cavities and substructures in the ISM around stars, the novelty of our data for the purpose of validating feedback models, and comparison of our findings with those in SN environment and SNR interaction studies.

\section{Observations and Methodology}\label{sec:obs}
We will carry out our analysis in the Local Group dwarf spiral galaxy, M33, which has the ideal combination of datasets for our purposes. At a distance of 0.84 kpc \citep{Freedman1991}, the star-forming dwarf with a stellar mass of $\sim$$3\times10^9$ \Msun \citep{vanderMarel2012} has a substantial atomic and molecular ISM visible at sub-cloud scale spatial resolution \citep[e.g.][]{Braun2012, Druard2014, Koch2018}. The galaxy has been the target of ground and space-based surveys of its stellar population and ISM for decades. We will use these published datasets to study the spatial correlation of evolved massive stars that will explode as SNe in future, with the density peaks in the ISM traced by the cold (atomic+molecular) gas at a resolution of $\sim$50 pc. The relevant ISM and stellar datasets, along with their motivation for usage, and measurement strategy, is described in the following sections.

\subsection{ISM maps} \label{sec:obs-ism}
\subsubsection{21 cm Atomic (\hi) Map} \label{sec:obs:21cm}
 
We use two \hi maps of M33, both obtained from the VLA and using short-spacing observations from the GBT \citep{Lockman2012}.
The coarser 20\arcsec\ (80~pc) resolution map is the 13-pt mosaic from \citet[][project ID: 14B-088]{Koch2018} using C configuration observations, which has a spectral resolution of $\sim$0.2 km s$^{-1}$ and rms noise of 2.8 K per channel. The data are feathered with the GBT \hi data from \citet{Lockman2012} with a single dish scaling factor of unity \citep{Koch2018}.

The second map uses new VLA B configuration observations (project ID: 17B-162) combined with the C configuration from \citet{Koch2018}.
We briefly describe these data here, and refer readers to further details in E. Koch et al. (in preparation).
We use the inner 7 pointings from the C-configuration observations and combine these with new B-configuration observations of the same pointings.
We image the data in 1.2~km s$^{-1}$ channels with natural weighting which yields a beam size of 8\arcsec (32~pc).
The rms noise per channel is 6.2 K per 1.2~km s$^{-1}$ channel.
We use the same feathering approach as described above with the C-configuration mosaic to incorporate the GBT \hi short-spacing information.

In this work, we use the integrated intensity (moment 0) maps calculated from both spectral-line data cubes following the signal masking approach from \cite{Koch2018}.
Briefly, we use a minimum $2\sigma$ threshold on the data, and keep only regions that contain a $>4\sigma$ peak and span 10~consecutive spectral channels, as appropriate for the typical \hi line widths sampled with $1$~\kms channel.
Following the findings from \cite{Koch2021}, we assume the \hi is generally optically thin on these spatial scales ($30\mbox{--}80$~pc) and convert to the atomic gas surface density in \Msunpc using
\begin{equation} \label{eq:h1}
    \frac{\Sigma_{\mathrm{HI}}}{\mathrm{M}_{\odot} \ \mathrm{pc}^{-2}} = 0.0196\ \mathrm{cos}(i) \left(\dfrac{I_{\mathrm{HI}}}{\mathrm{K\ km\ s^{-1}}}\right)\\
\end{equation}
where $i = 55^{\circ}$ \citep{Koch2018} is the inclination angle of M33.
The noise minimally varies across our field of interest, and so we adopt a median 3$\sigma$ completeness limit of 0.94 \Msunpc. The cubes also provided a map of the line-width of HI (km/s), which enters our calculations of the scale height and total gas volume density, as discussed in Appendix \ref{appendix:volumedensity}.

\subsubsection{ACA CO(2-1)-traced H$_2$ map} \label{sec:obs:aca}

We trace the molecular gas in M33 using ALMA Band 6 $^{12}$CO(2--1) observations obtained by the 7-m array in the Atacama Compact Array (ACA; project IDs: 2017.1.00901.S; 2019.1.01182.S).
The complete observations, including CO isotopologues, are explained in detail in E. Koch et al. (in preparation).
Briefly, the $^{12}$CO(2--1) observations are taken in 15 individually-observed mosaics that we image separately using a deconvolution approach similar to the PHANGS-ALMA imaging pipeline \citep{Leroy2021_pipeline}.
We then use linear mosaicking to combine each individually observed field into one complete mosaic covering the entire galaxy (see Figure \ref{fig:maps}).
Because of the northern location of M33 with respect to the ALMA site, the synthesized ACA beam is highly elongated in some observations (e.g., up to a factor of 3 between the major and minor axes).
The map we use here is convolved to the coarsest common round beam, giving us a final resolution of 12\arcsec ($\sim$49 pc).
We note that this matches the resolution of the IRAM 30-m CO(2--1) map \citep{Druard2014}, which maps a far larger region of M33. We verified that we recover consistent fluxes between the maps, demonstrating that no significant diffuse CO component is filtered out of the ACA data (Koch et al. in prep.).
However, the ACA data have a factor of 4 times finer spectral resolution (0.7~km s$^{-1}$) and a factor $\sim2$ higher sensitivity (25 mK per 0.7~km s$^{-1}$ channel).

We compute the CO(2--1) integrated intensity after calculating a signal mask with an approach similar to the ``strict'' method from \cite{Leroy2021_pipeline}, which we note does mildly differ from the approach used for the \hi but has been well-tested for CO emission that is clumpier with narrower line widths than the \hi.
As such, the signal masking approaches used for \hi and CO are appropriate given the expected properties (diffuse, extended vs. clumpy) of each tracer.
We convert the CO intensity to a mass surface density using,
\begin{equation} \label{eq:h2}
    \frac{\Sigma_{\mathrm{H_2}}}{\mathrm{M}_{\odot}\ \mathrm{pc}^{-2}} = \alpha_{\mathrm{CO}}R_{21}^{-1}\mathrm{cos}(i) \left(\dfrac{I_{\mathrm{CO}}}{\mathrm{K\ km\ s^{-1}}}\right)\\ 
\end{equation}
 where we assume $\alpha_{\mathrm{CO}} = 4.3\ \mathrm{M}_{\odot} \mathrm{pc}^{-2} (\mathrm{K\ km\ s^{-1}})^{-1} Z^{-1.6}$ is the empirical conversion factor between CO intensity and H$_2$ surface density, $Z$ is the metallicity normalized to the solar value,and $R_{21}=0.65$ is the CO (2-1)/(1-0) line-ratio \citep{Sun2018} . We use a constant $\alpha_{\mathrm{CO}}$ prescription for simplicity, assuming $Z = 0.6$ Z$_{\odot}$, appropriate for the inner 5 kpc of M33 \citep{Rogers2022}\footnote{Since $\alpha_{\mathrm{CO}}$ can vary with metallicity, we also checked our results with a variable $\alpha_{\mathrm{CO}}$ prescription, namely one where $Z$ varies with galactocentric radius based on the measured metallicity gradient in M33 \citep{Bresolin2011}. The difference in $\Sigma_{\mathrm{H_2}}$ between our variable and constant $\alpha_{\mathrm{CO}}$ is $\sim$30-40\%, much smaller than the orders of magnitude variation in $I_{\mathrm{CO}}$ and $I_{\mathrm{HI}}$.}. From our noise cubes, we obtain a median 3$\sigma$ noise (assuming detection over 3 channels) of 0.16 K km s$^{-1}$, and therefore a median completeness limit of our H$_2$ maps of 1.37 \Msunpc.

\subsubsection{High-resolution ALMA CO(2--1)-traced H$_2$ map} \label{sec:highresalma}

We also make use of a single ALMA field observed around a WR and SNR in close proximity (Figure \ref{fig:maps}, box region) to highlight the finer substructure in cloud environments of SNe that can be revealed at much higher resolution. Specifically, we look at the region around SNR L10-045 from ALMA Band 6 data published in \citet[][project ID: 2018.1.00378.S]{Sano2021b}. The area observed is a single ALMA 12-m pointing. These data resolve 0.46\arcsec ($\sim2$~pc) scales in $^{12}$CO(2-1) and so resolve the filamentary structure within this single GMC. The data we show here are re-processed, using the ALMA delivered QA2 data from the archive with imaging using the PHANGS-ALMA pipeline \citep{Leroy2021_pipeline}. We image the data using Brigg's weighting with \texttt{robust=0.5} in $1.2$~km s$^{-1}$ channels. The corresponding rms is 38 mK per $1.2$~km s$^{-1}$ channel. We signal mask and generate moment maps using the PHANGS-ALMA pipeline \citep{Leroy2021_pipeline}; specifically we use the integrated intensity map with strict signal masking.

\subsection{Stellar Catalogs} \label{sec:obs-stars-wr}
To trace the locations where stars will explode in M33, we use catalogs of WRs, RSGs and SNRs. We briefly describe the motivation for using such evolved massive stars as tracers of future supernovae in Section \ref{sec:obs-why}, and details of the catalogs themselves in the subsequent subsections.
\subsubsection{Motivation for using evolved massive stars}\label{sec:obs-why}
We use evolved massive stars (WRs and RSGs) since they have been observationally confirmed as SN (or potential SN) progenitors. Of the two, RSGs have been directly identified as progenitors of Type II SNe, which form the majority category of core-collapse SNe in flux-limited surveys \citep{Smartt2009, Smartt2015}. WRs, characterized by strong winds and stripped stellar envelopes, have been classically considered progenitors of stripped-envelope (SE) SNe, which form the second dominant category of core-collapse (CC) SNe after Type II \citep[e.g.]{Smartt2009}, although this has been called into question owing to several inconsistencies in the properties of WRs and SE-SNe \citep[e.g.][]{Eldridge2013, Drout2011, Taddia2015, Smith2011, Graur2017}\footnote{A small fraction of SE-SNe however, may still be consistent with WR-like progenitors, particularly those associated with long-gamma-ray bursts \citep[e.g. see][and references therein]{Crowther2007, Smartt2015, Roy2021} and strong circumstellar interaction \citep[e.g.][]{VanDyk2018, Xiang2019, Horesh2020, GalYam2022, Kool2021, Pellegrino2022}}. To first order, we assert that our WRs at least trace the locations of the most massive stars in M33, since WRs are expected to be $>$40 \Msun ($>$20 \Msun) based on single (binary) evolution models \citep{Eldridge2017}. Additionally, WRs and SE-SNe appear to evolve in environments with very similar star-formation rates \citep[e.g.,][]{Kangas2017, Maund2018}, so we expect the explosion sites of WRs in the ISM to be similar to that of SE-SNe.

Another major reason for using RSG and WRs is because of their significantly shorter lifetimes than their main-sequence (OB) counterparts, implying that they are spatially and temporarily `near' the location of their future SN explosion sites. The typical durations of OB stars are about 10 times longer than the post-main-sequence phases where RSGs and WRs manifest \citep{Ekstrom2012}. As a result, for the same velocities, OB stars would travel 10 times farther before exploding, so their explosion sites have a larger `error' circle from the present location, compared to RSGs and WRs (for reference, a star moving at a velocity $v_*$ will move $(\mathrm{10pc}) (v_*/10 \mathrm{km/s}) (t_*/1 \mathrm{Myr})$.

\subsubsection{Wolf-Rayet Stars}
We use the M33 WR catalog of \cite{Neugent2011}. The study cataloged 206 WRs within 0.32 deg$^2$ ($\sim$ 8.5 kpc radius) of M33 using observations from the 4\,m Kitt Peak Mayall telescope and Mosaic CCD. Candidate WRs were identified with crowded-field photometry and image subtraction with narrowband filter images centered on \ion{C}{3} $\lambda$4650 and \ion{He}{2} $\lambda 4686$ lines (typically seen in WRs), and then spectroscopically confirmed using the 6.5 m MMT Hectospec instrument. 
Objects were classified as WN (strong lines of He, N) and WC (strong lines of He, C), with further subclassification (e.g. WN3-WN8) based on relative line strengths of He, C, N and O ionization states as per convention \citep{Conti1988}. 

The WR sample is expected to have high completeness according to \cite{Neugent2011}, who estimated a 95\% completeness based on comparisons with previous studies in the Local Group \citep{MasseyJohnson98,Neugent2018}.  \cite{Neugent2023} further confirmed the high completeness of the M33 WRs; the faintest WR star in the catalog is still about $\sim$1 mag brighter than their nominal sensitivity limit. We also note from calculations in Section \ref{sec:results-spatial} that even in regions like NGC 604 that are expected to have high extinction, the observed number of WRs are consistent with expectations from the underlying star-formation history in the region \citep{Lazzarini2022}. The MMT spectroscopy also ensured high sample purity, removing cases where previously-identified WRs in NGC 604 and NGC 595 \citep{Drissen2008} were revealed to be O-type stars with WR-like emission lines from the spectroscopy of \cite{Neugent2011}. 

Of the 206 WRs, we restrict our catalog to 119 objects within the footprint of our CO data (described in Section \ref{sec:obs-ism}). 

\subsubsection{Red Supergiant Stars} \label{sec:obs-stars-rsg}
We use the RSG catalog constructed by \cite{Massey2021} from a UKIRT $J-K_s$ band survey of the inner 3 deg$^2$ of M33. Foreground stars were separated from M33 stars using \emph{Gaia} proper motion and parallax measurements, and RSGs were separated from asymptotic giant branch (AGB) stars and background galaxies using cuts in $J-K_s$ colorspace, constructed to optimize RSG selection based on previous tests in the Magellanic Clouds \citep{Yang2019, Neugent2020}. The final catalog consists of 7088 RSGs, of which we use the 812 RSGs inside our CO footprint. Physical quantities like luminosity and temperature were also reported by \cite{Massey2021} after correcting for extinction and reddening, and appear to be in qualitative agreement with predictions of stellar evolution models \citep{Ekstrom2012}.

\subsubsection{Supernova Remnants} \label{sec:obs-stars-snrs}
As an interesting check, we will also compare densities at the locations of WRs and RSG catalogs with sites of historical SN explosions (in the last 10$^{4}$-10$^{5}$ yrs) in M33 using the known SNR population. We use the SNR sample in M33 compiled in \cite{White2019}, which is the latest and most complete sample of SNRs in the galaxy. The SNRs were discovered using optical-line emission data, specifically those with [\ion{S}{2}]/H$\alpha \gtrsim 4$ (to separate from photoionized nebulae) using data from 4m Mayall Telescope and 0.9m Burrell Schmidt telescope at KPNO ratios \citep{Long2010, Lee2014}. These candidates were further spectroscopically confirmed as SNRs with 6.5 MMT \citep{Long2018}, Chandra X-ray observations \citep{Long2010} and sensitive JVLA observations at 1.4 and 5 GHz taken in \cite{White2019}, covering roughly the inner 0.85 deg$^2$ (0.35 deg$^2$ at 5 GHz) of M33. Of the 215 SNRs in the sample, we use the 71 inside the footprint of the CO map.

\begin{figure*}
    \centering
    \includegraphics[width=0.94\textwidth]{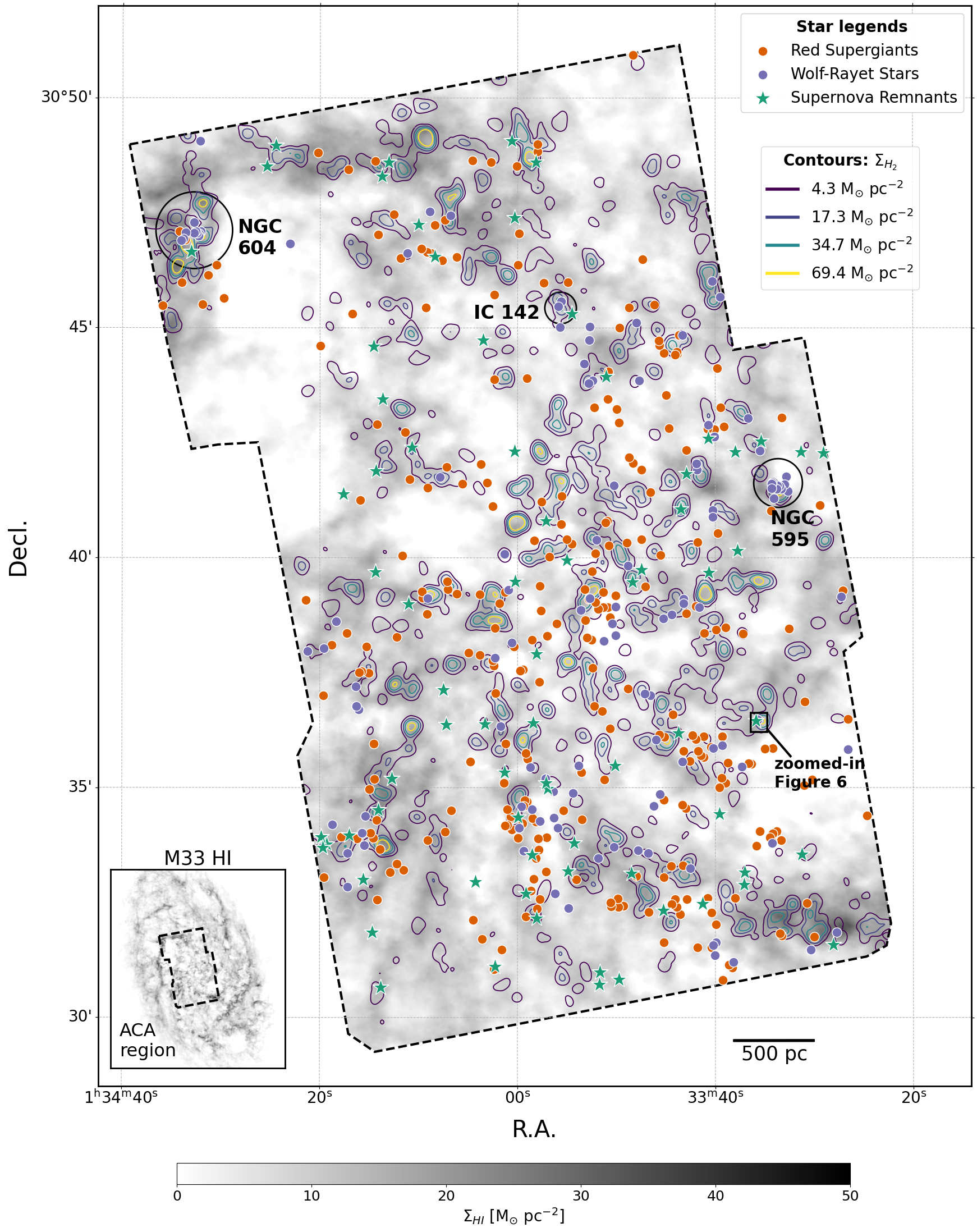}
 \caption{The spatial distribution of red supergiants (RSGs), Wolf-Rayet stars (WRs) and supernova remnants (SNRs) with respect to the cold (H$_2$ + \hi) ISM in the M33 ACA survey region (shown in the inset plot). Orange circles represent RSGs with log $(L/L_{\odot})\geq 4.7$ (roughly corresponding to $\gtrsim$12\Msun). Bluish circles show WRs and green stars show SNRs. \hi is shown in grey-scale with the density range shown in the bottom colorbar. Contours of CO (2-1) emission (converted to $H_2$ surface densities) are also shown at levels of 1.9, 7.6, 15.3 and 30.6 M$_{\odot}$ pc$^{-2}$. The prominent star-forming regions in the ACA footprint -- NGC 604, IC 142 and NGC 595 -- are specifically labelled (showing high WR concentrations). The black box shows a region containing a WR and SNR with higher-resolution ALMA data in Figure 6. The map shows the 50 pc-scale cold ISM environments where core-collapse SNe will occur in the future (or have occured in the case of SNRs). Statistical analysis of these environments and their implications are discussed in Section \ref{sec:results} and \ref{sec:discussion}.}
    \label{fig:maps}
\end{figure*}

\subsubsection{Recording ISM densities around stars}
With the density maps generated in accordance with Section \ref{sec:obs:21cm} and \ref{sec:obs:aca}, we record the surface densities of \hi and H$_2$ (derived from Eqs \ref{eq:h1} and \ref{eq:h2}) at the pixel location of each star in our WR, RSG and SNR catalog, effectively representing the projected density averaged over a 50 pc diameter region (roughly the PSF of the maps), similar to the methodology in MC22. Our measured ambient densities on 50 pc scales should also be representative of the densities at the time of explosion in most regions, since the turbulent sound-crossing timescale, for a typical $\sigma = 10$ km/s, is about 5 Myrs, longer than the expected time left for the evolved massive stars to explode \citep[$<$0.5 Myrs for stars $>$30 \Msun, and $\sim$0.5-4 Myrs for 8-30 \Msun RSGs,][]{Ekstrom2012}.\footnote{SN blastwaves however can alter the gas distribution on a much shorter timescale due to their faster speeds. We return to this point in the discussion in Section \ref{sec:densityage}.}

\subsection{Galaxy simulation} \label{sec:data:sims}
In Section \ref{sec:disc:sims} we will compare our measured ambient densities (cm$^{-3}$) around WRs and RSGs with the ambient densities of SNe in a hydrodynamical simulation of a dwarf starburst galaxy system in \cite{Lahen2020}. The goal is to show that our observations of SN environments provide a simple line of comparison with hydrodynamical simulation of galaxies for validation of the underlying stellar feedback models. We pursue a preliminary comparison in this paper with \cite{Lahen2020} as a demonstration of how such comparisons can take place, with more detailed investigations to be done in future work.

We provide here a brief summary of the simulation, and refer the reader to the paper for details. The simulation follows the merger of two identical gas-rich dwarf galaxies of initial stellar masses of $2\times 10^7$ \Msun, at a baryonic mass resolution of \mbox{$\sim4$ \Msun} and sub-pc spatial resolution. We chose this simulation for our example because the high baryonic resolution ($\sim$4 \Msun) leads to well-resolved ISM environments \citep{Steinwandel2020}, the starburst system results in a wide range of ISM densities and star formation rates, and because each SN event can be directly traced back to a single progenitor star. The simulation was performed using \textsc{SPHGal} (\citealt[][and references therein]{Hu2017}), an improved version of the smoothed particle hydrodynamics code \textsc{GADGET-3} \citep{Springel2005}, which uses a simplified chemical network to model the non-equilibrium cooling and heating processes within the ISM including H$_2$ formation and carbon chemistry. Stellar feedback processes driving the multiphase ISM include the interstellar far-ultraviolet radiation field, photoionizing radiation, core-collapse supernovae and winds of asymptotic giant branch stars, all modelled according to the stochastically sampled IMF (for details see \citealt{Hu2017, Lahen2020}). Massive stars (\mbox{$>4$ \Msun}) are realized as single particles in the simulation, allowing us to measure the densities around each SN progenitor right before the explosion. 

We will compare both the ``local'' and ``average'' ambient densities of SNe in the simulation with the observations. Here ``local'' refers to the density immediately surrounding the SN particles, which we obtain from the weighted average density of the 100 closest gas particles. However, the local densities of stars in observations are not directly measurable. To better compare the simulation with observations, we obtain the column-averaged densities by projecting the gas distribution in each simulation snapshot onto a grid along the z-axis, at a resolution of 50 pc per pixel similar to our observation, and then dividing the surface densities by a scale height of 100 pc. The gas density maps are recorded in 1 Myr intervals, and the SN densities are measured in the pixels that correspond to the locations of the evolved massive stars (\mbox{$8$--$50$ \Msun}) in the snapshot prior to the SN explosion, i.e. up to 1 Myr earlier. We record densities for all SN progenitors produced during the pre-starburst phase spanning 70-140 Myrs in the simulation when the two galaxies are relatively quiescent and consistent with the specific star-formation rate of M33 (see Fig. 5 in \citealt{Lahen2020}). We expect this phase to be a better match to M33, which is currently not undergoing a starburst. We divide these SN progenitors into those with initial stellar mass of \mbox{8-30 \Msun} and \mbox{$>$30 \Msun} to directly compare with the RSG and WR stars. 

\section{ANALYSIS} \label{sec:results}
With the datasets described in Section \ref{sec:obs}, we quantify differences in the ISM properties of the different populations by comparing the spatial distributions of RSGs, WRs and SNRs with respect to the \hi and H$_2$ distribution within the ACA area (Figure \ref{fig:maps}). Surface densities of H$_2$ and \hi at the pixel locations of stars are summarized in Table \ref{tab:densities} and shown in Figures \ref{fig:3paneldensities} and \ref{fig:3paneldensities_rsgs}.

\subsection{Spatial distribution of stars and gas} \label{sec:results-spatial}
The relative distribution of the cold (H$_2$ + \hi) ISM and stars at $\sim$50 pc spatial resolution is shown in detail in Figure \ref{fig:maps}, and the corresponding density ranges are summarized in Table \ref{tab:densities}. Locations of the stars and SNRs are shown as colored circles/stars, and the \hi and H$_2$ distributions are shown as greyscale and contours (respectively) in units of surface density. 

The \hi distribution fills most of the survey area in Figure \ref{fig:maps} compared to H$_2$, and shows the characteristic frothy structure typically seen in nearby galaxies \citep[e.g][]{Stanimirovic1999,Walter2008,Pingel2022}. In comparison, the H$_2$ contours are clumpier, coinciding with dense molecular clouds in active star-forming regions. The detailed structure of the \hi and H$_2$ of M33 will be covered in a forthcoming publication (Koch et al, in prep). We note that M33 is about an order-of-magnitude lower in stellar mass and average molecular surface density compared to the population of nearby, star-forming galaxies in PHANGS analyzed by MC22 \citep{Leroy2021}. Our work is therefore complementary to MC22 in that it probes SN environment densities in a lower-mass dwarf spiral galaxy. 

While the stars generally follow the gas distribution, we note the stark contrast in the relative spatial distribution of the gas and WRs, compared to SNRs and RSGs. Dense concentrations of WRs in Figure \ref{fig:maps} are particularly prominent in major star-forming regions such as NGC 604, NGC 595, and IC 142, as well as the numerous \hii regions interspersed in the southern arm. This WR distribution is a result of the abundant young ($<$10 Myr) stellar populations in these regions. For example, the total stellar mass formed in the last $\sim$10 Myrs in NGC 604 is $\left(1.06_{-0.42}^{+0.67}\right) \times 10^5$ \Msun\ based on the PHATTER survey \citep[][note that maps measure star formation histories in $\sim$100-pc-sized cells]{Lazzarini2022}. The production rate of WRs is roughly 1.3 WR per 10$^4$ \Msun  \citep{BPASS, DornW2018}, which gives an expected number of WRs in NGC 604 to be $14^{+8}_{-6}$, consistent with the 11 observed in the region. In contrast, the 10 Myr stellar mass of IC 142 is about 7 times lower, so only about 2 WRs are expected in the region, consistent with Figure \ref{fig:maps}. NGC 595 on the other hand has the same number of WRs as NGC 604 but with a factor of 2 smaller stellar mass, likely due to a more complex dependence on the underlying metallicity and star-formation history \citep{Drissen1993}. This tendency of WRs to concentrate in actively star-forming regions is likely due to their higher zero-age main-sequence masses ($>$30 \Msun) and leads to a stronger correlation of the WR population with dense molecular gas compared to RSGs ($\gtrsim$8-30 \Msun) and SNRs, which we discuss in more detail in Section \ref{subsec:snrs}.

\subsection{ISM densities at the location of stars} \label{sec:physcorr}
\begin{figure*}
    \includegraphics[width=\textwidth]{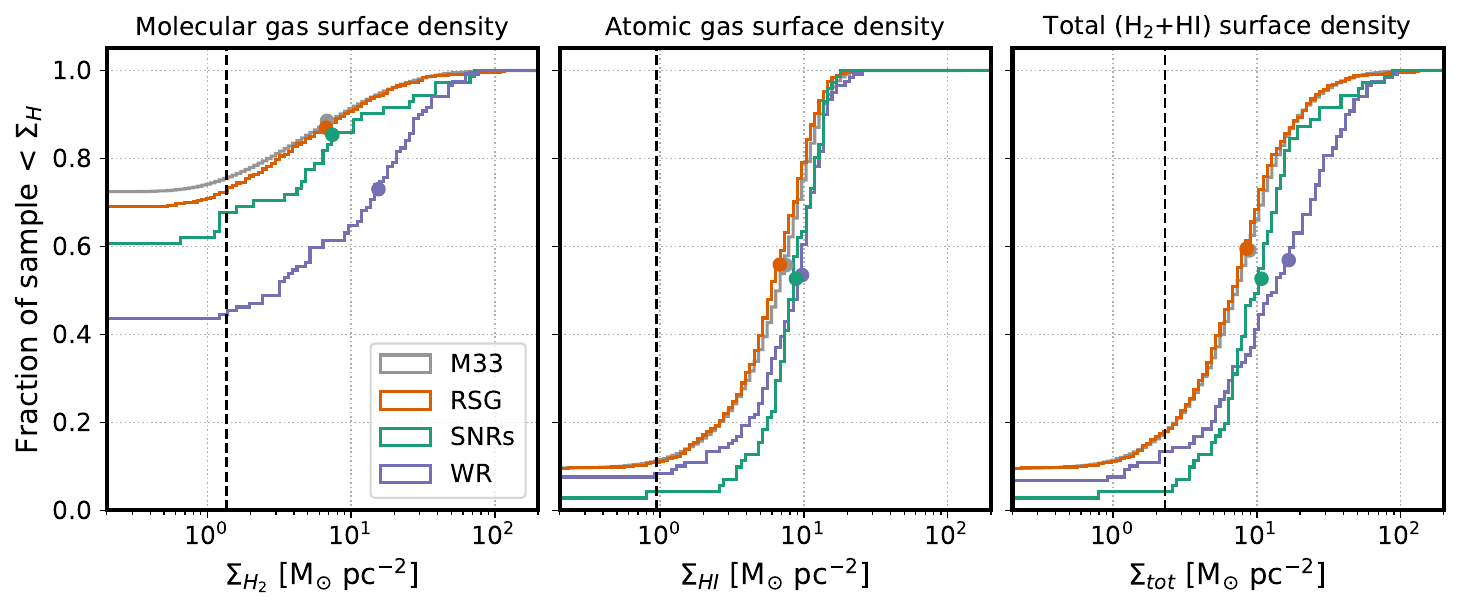}
    \caption{Cumulative distribution functions showing H$_2$, \hi, and total (H$_2$ + \hi) surface densities. Distributions are shown for the locations of WRs (blue lines), SNRs (green lines) and RSGs (red lines), while grey histograms denote the distribution of densities of the full M33 region, representing what would be expected for a random distribution. Vertical dashed lines denote the completeness limit of each dataset. Colored circles show the median of the distribution above the completeness limit. The distributions show a higher probability of more massive stars (e.g. the WRs) to evolve in denser gas.}
    \label{fig:3paneldensities}
\end{figure*}

We now compare the surface density distributions\footnote{The density measurements will be made publicly available after acceptance and publication of the manuscript. Please feel free to contact us if you would like to use the dataset beforehand.} of gas tracers at the locations of different categories of stars in Figure \ref{fig:3paneldensities}. We also list their median, 16th-84th percentile range (corresponding to a 1$\sigma$ interval), and fraction of non-detections (i.e., pixel values below the gas surface density completeness limit) in Table \ref{tab:densities}. The median and percentiles are estimated only using values above the completeness limits.

\begin{figure*}
    \includegraphics[width=\textwidth]{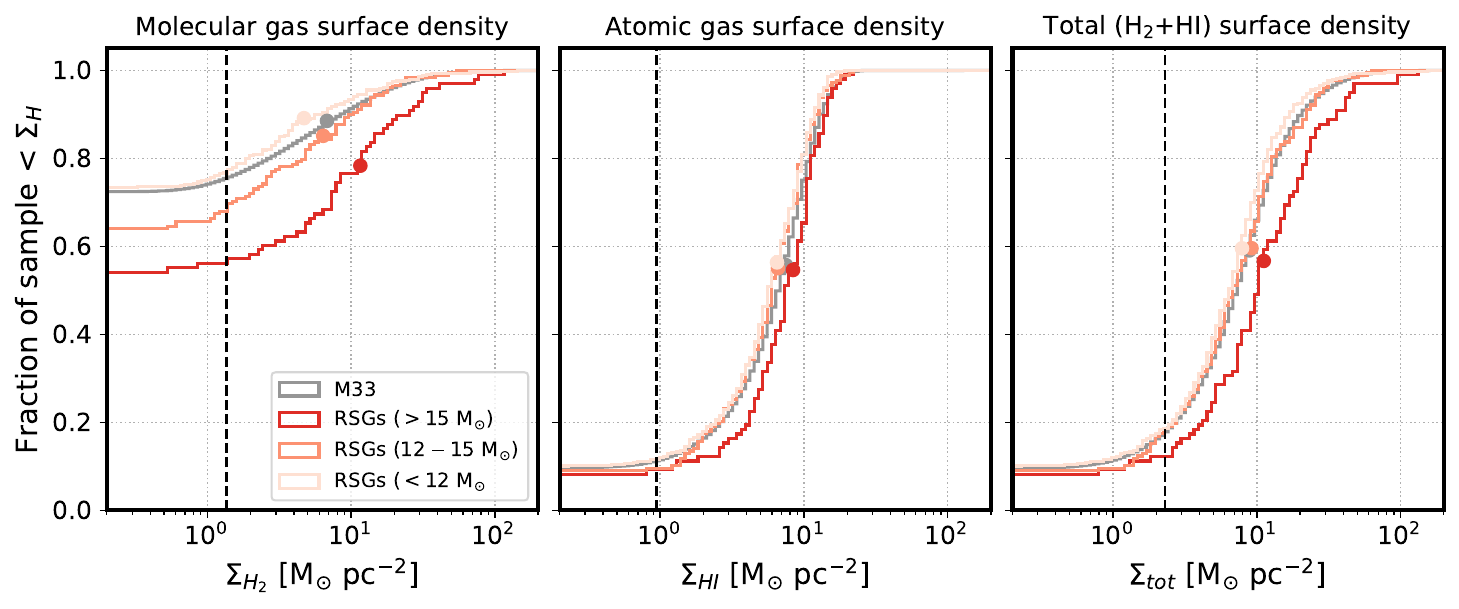}
    \caption{Cumulative distribution functions showing H$_2$, \hi and total gas surface densities (as in Figure \ref{fig:3paneldensities}). Here only the RSGs are plotted, dividing them into 3 luminosity bins, roughly tracing mass ranges of 8-12 \Msun (peach lines), 12-15 \Msun (orange lines) and $>$15 \Msun (red lines). Grey histograms denote the distributions for the full M33 region. Colored circles show the median of the distribution above the completeness limit.}
    \label{fig:3paneldensities_rsgs}
\end{figure*}

To visualize the distributions, we plot the cumulative distribution function (CDF) of H$_2$ and \hi surface densities for WRs, RSGs and SNRs in Figures \ref{fig:3paneldensities} and \ref{fig:3paneldensities_rsgs}. The CDF captures the same information as a standard histogram or kernel density estimation without the need to adopt bins or a smoothing kernel. It allows easy comparison between our samples, which have very different overall sample size. The CDFs flatten to 1 at the highest pixel value of the maps, i.e., 100\% of the pixels are below that maximum value. On the other end, the CDFs flatten to a limiting fraction value as they reach the completeness limits of the maps (vertical dashed lines). One should read this limiting value as the fraction of that object sample without a detection in the associated gas type. This limiting value is quite high for the H$_2$ distribution of M33 ($\sim75\%$), reflecting the fact that the molecular gas is concentrated into individual dense clouds, and as a result CO emission covers only a small fraction of the total area in M33 (Figure \ref{fig:maps}). On the other hand, atomic (\hi) gas has a much larger filling fraction, and nearly 90\% of the pixels have some \hi detection. 

Notably, we find WRs coincident with H$_2$ column densities on average 3 times higher than that of all pixels in M33 and those containing RSGs, and almost 2 times higher than pixels with SNRs (Table~\ref{tab:densities}).  Only 4\% of the RSG pixels and 8.5\% SNR pixels have densities $\geq20$ \Msunpc \citep[typical for molecular clouds in M33,][]{Rosolowsky2007} while 21\% of the WR pixels are above that value. Also, only 45\% of the WRs are located in H$_2$ column densities below our detection limit (i.e. $<1.4$ \Msunpc), compared to 68\% and 72\% for RSGs and SNRs respectively (we will frequently refer to this as the `non-detection' fraction). 

\begin{deluxetable*}{lrccccccccc}
\tablecaption{Surface densities of detectable H$_2$ and \hi around different categories of stars, and the corresponding fraction of stars with H$_2$/HI non-detections, as measured in Section \ref{sec:physcorr}. The first row are values for all pixels in M33, and subsequent rows denote pixel values of stars in the different categories considered in the paper. Columns from left to right: 
Category of star, 
sample size ($N$), the median gas surface density ($\langle \Sigma \rangle_{\mathrm{med}}$), the 16th-84th percentile range of surface densities ($\langle \Sigma \rangle_{\mathrm{16-84}}$), and non-detection fractions ($f^{\mathrm{non}}$) of H$_2$, \hi and the total (H$_2$ + \hi) gas column respectively. All median and percentile calculations are calculated for values higher than the respective completeness limits. \label{tab:densities}}
\setlength{\tabcolsep}{4pt}
\tablehead{
\colhead{Category} & 
\colhead{$N$} & 
\colhead{$\langle\Sigma_{\mathrm{H_2}}\rangle_{\mathrm{med}}$} & 
\colhead{$\langle\Sigma_{\mathrm{H_2}}\rangle_{\mathrm{16-84}}$} & 
\colhead{$f^{\mathrm{non}}_{\mathrm{H_2}}$} &
\colhead{$\langle\Sigma_{\mathrm{HI}}\rangle_{\mathrm{med}}$} & 
\colhead{$\langle\Sigma_{\mathrm{HI}}\rangle_{\mathrm{16-84}}$} & 
\colhead{$f^{\mathrm{non}}_{\mathrm{HI}}$} & 
\colhead{$\langle\Sigma_{\mathrm{tot}}\rangle_{\mathrm{med}}$} & 
\colhead{$\langle\Sigma_{\mathrm{tot}}\rangle_{\mathrm{16-84}}$} & 
\colhead{$f^{\mathrm{non}}_{\mathrm{tot}}$} \\
\colhead{} & 
\colhead{} & 
\colhead{(\Msunpc)} & 
\colhead{(\Msunpc)} & 
\colhead{} & 
\colhead{(\Msunpc)} & 
\colhead{(\Msunpc)} & 
\colhead{} & 
\colhead{(\Msunpc)} & 
\colhead{(\Msunpc)} & 
\colhead{}
}
 \startdata
M33 & -- & 6.8 & 2.5--20.5 & 0.75 & 7.5 & 3.6--12.4 & 0.11 & 8.9 & 4.5--18.0 & 0.17 \\
WRs & 119 & 15.5 & 3.8--33.8 & 0.45 & 9.7 & 5.0--14.0 & 0.08 & 16.7 & 6.5--39.8 & 0.13 \\
RSGs (all) & 812 & 6.7 & 2.4--18.9 & 0.72 & 6.8 & 3.5--11.4 & 0.11 & 8.5 & 4.5--17.7 & 0.18 \\
\textbf{RSGs ($>15$ M$_{\odot}$)} & 98 & 11.6 & 4.8--30.5 & 0.56 & 8.4 & 4.8--13.5 & 0.09 & 11.2 & 5.4--26.2 & 0.12 \\
\textbf{RSGs ($12-15$ M$_{\odot}$)} & 178 & 6.4 & 2.4--17.5 & 0.68 & 6.6 & 3.2--11.3 & 0.1 & 9.2 & 4.3--18.9 & 0.18 \\
\textbf{RSGs ($<12$ M$_{\odot}$)} & 536 & 4.7 & 2.1--18.2 & 0.77 & 6.5 & 3.4--11.1 & 0.12 & 7.9 & 4.3--15.3 & 0.18 \\
SNRs & 71 & 7.4 & 4.5--33.8 & 0.68 & 8.8 & 5.9--13.8 & 0.04 & 10.8 & 6.4--18.3 & 0.04 \\
\enddata
\end{deluxetable*}

These results suggest that WRs are much more closely associated with dense gas than the bulk RSG and SNR population, which is consistent with the idea that WRs arise from more massive progenitors that have shorter lifetimes, and thus more likely to be found near their birth clouds. A significant fraction of WRs however ($\sim$45\%), appear outside any detectable molecular gas, which has important implications for feedback that will be discussed later in Section \ref{sec:discussion}.

Unlike H$_2$, most of our stars are evolving in pixels with some detectable atomic gas. The majority of the pixels in our ACA region have \hi detections down to densities $\sim$$0.94$ \Msunpc. Of the 45\% of WRs (53) without a significant H$_2$ detection, 83\% (44 out of 53) are in pixels with detectable atomic gas. The statistics are also similar for RSGs and SNRs: about 85\% RSGs with H$_2$ non-detections (499 out of 587) and 94\% SNRs with H$_2$ non-detections (45 out of 48) are in pixels with detectable atomic gas. 
About $\sim$10\% of the stars in all categories are \hi non-detections, but as shown with more sensitive low-resolution \hi data from \cite{Koch2018} in Appendix \ref{sec:Cconfig}, there is still some atomic gas at the locations of these stars, just with surface densities below our sensitivity limit of $\Sigma_{HI}\lesssim 1$ \Msunpc. 

The resulting \emph{total} surface density of neutral gas ($\Sigma_{HI} + \Sigma_{H_2}$), which would be of interest in simulations, is shown in the bottom left panel of Figure \ref{fig:3paneldensities}, clearly showing a mix of the \hi and H$_2$ distribution properties. Almost 80-90\% of stars in the three categories are evolving in column densities $\gtrsim$2 \Msunpc, but WRs are still found in environments almost twice as dense as RSGs and SNRs that have lower-mass progenitors than WRs. 

The correlation between progenitor mass and ISM densities is more explicity shown in Figure \ref{fig:3paneldensities_rsgs}. Here again we show the \hi and H$_2$ CDFs similar to Figure \ref{fig:3paneldensities} but only for RSGs, divided into three luminosity (and therefore roughly progenitor mass) bins: $\mathrm{log}(L/L_{\odot}) = 4.2-4.7$, $\mathrm{log}(L/L_{\odot}) = 4.7{-}5$, and $\mathrm{log}(L/L_{\odot})>5$. Based on comparison with stellar evolution models shown in Figure 8 of \cite{Massey2021}, these ranges roughly correspond to zero-age main-sequence masses of $8-12$\Msun, 12--15\Msun, and $>$15\Msun. From here on, we will refer to these RSGs by their mass ranges. The lower limit of $\mathrm{log}(L/L_{\odot})=4.2$ also helps to avoid contamination from bright AGB stars. We find that, similar to the results in Figure \ref{fig:3paneldensities}, a greater fraction of higher mass RSGs are evolving in denser H$_2$ and \hi gas, with the differences in \hi gas distributions being less pronounced than H$_2$. Notably, only 56\% of RSGs $>15$\Msun\ have H$_2$ non-detections, compared to the 77\% of $8-12$\Msun\ RSGs having H$_2$ non-detections. This is similar to how WRs have a smaller fraction of H$_2$ non-detections than the full RSG sample.

\subsubsection{Statistical significance of the distributions} \label{sec:stats}
Figures \ref{fig:3paneldensities} and \ref{fig:3paneldensities_rsgs} suggest systematic differences between the gas distributions found at the locations of different evolved stars. But are these differences statistically significant?

We first check this with the two-sample Kolmogorov-Smirnov (KS) Statistic, which tests the similarity between two unequal-sized distributions by measuring the maximum distance between the sample CDFs. While the KS test has weak sensitivity to differences at the tails of distributions, it is quite sensitive to the difference in medians. We compare the \hi and H$_2$ surface density CDFs of WRs, the three subcategories of RSGs, and SNRs with that of the bulk H$_2$ and \hi distribution of M33 using the KS test. We implement this using \texttt{scipy.stats} package, and use the two-sided $p$-value returned by the test to assess whether the null hypothesis, that the locations of massive stars are drawn from the bulk H$_2$ and \hi distribution, can be rejected. We use the standard threshold of $p<0.05$ to reject the null hypothesis. The KS-statistic and $p$-values are listed in Table \ref{tab:stats}.
\begin{deluxetable}{lcrcr}
\tabletypesize{\small}
\setlength{\tabcolsep}{1.3pt}
\tablewidth{-1in}
\tablehead{
\colhead{\hspace{-1cm}Category} &
\colhead{KS (H$_2$)} & 
\colhead{$p$} & 
\colhead{KS (\hi)} &
\colhead{$p$}
}
\tablecaption{Summary of KS statistic tests for different star categories, specifically comparing their H$_2$ and \hi distributions with the bulk distribution of M33 as described in Section \ref{sec:stats}. Columns from left to right: Name of stellar category, KS Statistic for H$_2$ surface density distribution, $p$-value for H$_2$ KS test, KS Statistic for \hi surface density distribution, and $p$-value for \hi KS test. Boldfaced numbers emphasize $p$-values less than or equal to 0.05. \label{tab:stats}}
\startdata
WRs & 0.33 & $\mathbf{9.2 \times 10^{-12}}$ & 0.20 & $\mathbf{1.6\times10^{-4}}$ \\
RSGs (all) & 0.03 & 0.37 & 0.06 & $\mathbf{3.4\times10^{-3}}$ \\
RSGs & & & \\
-- ($>15$ M$_{\odot}$) & 0.22 & $\mathbf{1.5\times10^{-4}}$ & 0.11 & 0.17 \\
-- ($12-15$ M$_{\odot}$) & 0.08 & 0.24 & 0.08 & 0.15 \\
-- ($<12$ M$_{\odot}$) & 0.04 & 0.22 & 0.09 & $\mathbf{4.3\times10^{-4}}$ \\
SNRs & 0.12 & 0.24 & 0.25 & $\mathbf{3.0 \times 10^{-4}}$ \\
\enddata
\end{deluxetable}

The results are summarized in Table \ref{tab:stats}. For H$_2$, we find that the WRs have the largest difference between its CDF and the bulk M33 H$_2$, and the null hypothesis can be rejected with high confidence, at a significance level $
p\ll$0.05. A similar result is also found for the RSGs $>$15 \Msun, but not for RSGs $<$15 \Msun\ and SNRs, which appear at least marginally consistent with having been drawn from the same parent population as the overall distribution within our field of view. For \hi, the WRs and SNRs have the largest CDF difference, and the null hypothesis is rejected at the 0.05 level. The \hi CDF of high-mass RSGs, similar to H$_2$, have the largest difference of the three RSG categories with the bulk distribution, but it is not statistically significant in this case. The \hi of low-mass RSGs on the other hand appear to have a statistically significant difference, but smaller than the high-mass ones. A close inspection of Figure \ref{fig:3paneldensities_rsgs} shows that the low-mass RSGs are exploding in slightly lower densities than both the bulk \hi and H$_2$ of M33.  We also re-checked this with the Anderson-Darling (AD) test, which is more sensitive to the tails of the distributions. We find that the above cases where $p<0.05$ with the KS test also have $p<0.05$ with the AD test.
The AD test returns two cases implying statistical significance at $p<0.05$ where the KS test does not: the H$_2$ CDFs of the 12-15 \Msun RSGs, and SNRs.

These results provide an initial quantitative support (which we  investigate further in the next Section \ref{sec:sfrtracers}) that higher mass progenitors (as traced by WRs and RSGs $>$15 \Msun) are indeed evolving in denser H$_2$ pixels than the lower mass stars (RSGs $<$15 \Msun), and their densities are not just random sampling from the bulk distribution of M33. The same holds for the \hi densities of WRs, though the difference between the \hi densities of RSGs $>15$\Msun and M33 is not statistically significant. Visually, these results relate back to the spatial distributions in Figure \ref{fig:maps}, where we see WRs appear more frequently within H$_2$ clouds marked by the contours, and sometimes well within giant molecular clouds like in NGC 604 and NGC 595, compared to the RSGs.

\subsection{Comparison with the general stellar population} \label{sec:sfrtracers}
\begin{figure*}
    \centering
    \includegraphics[width=0.9\textwidth]{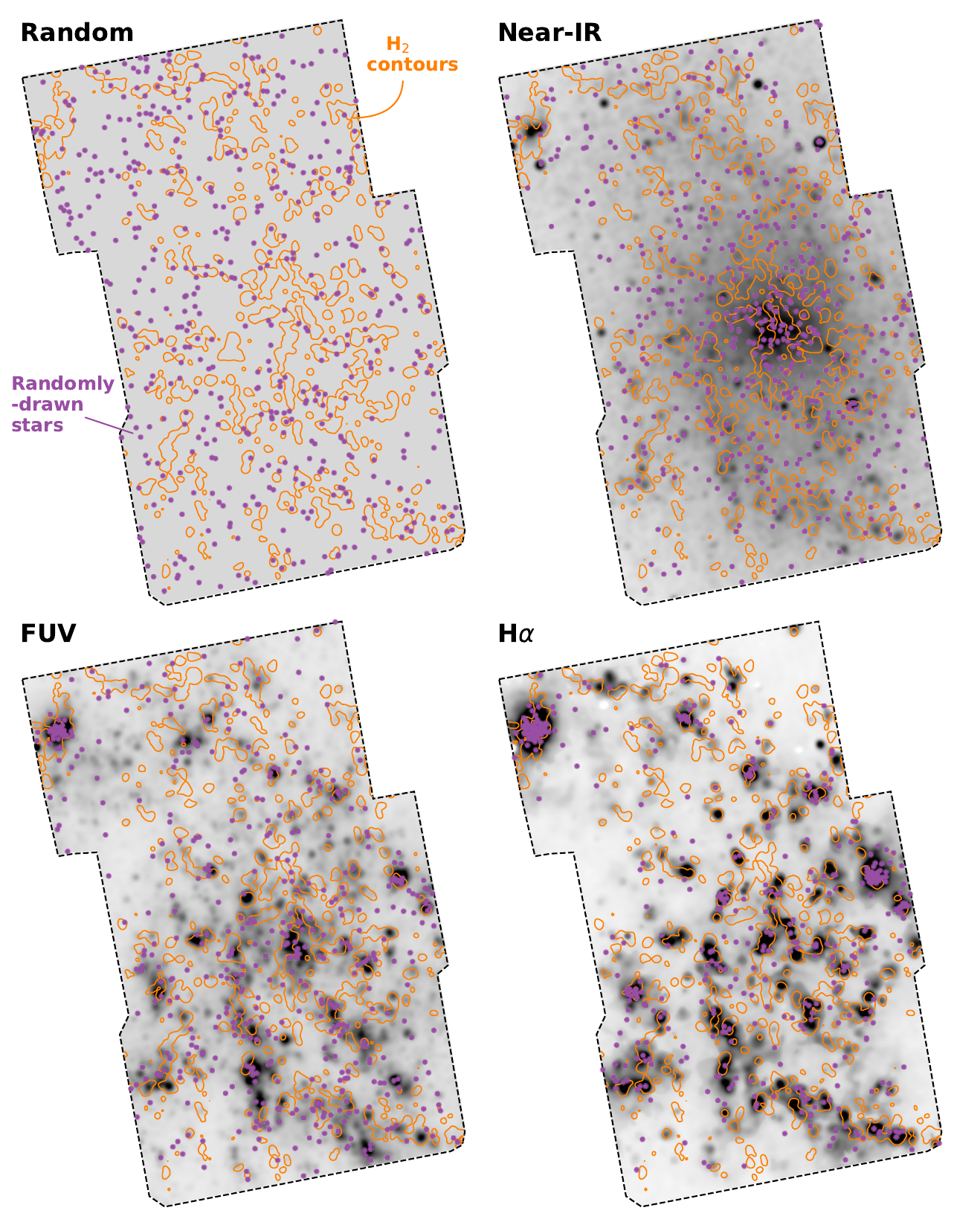}
    \caption{Distribution of mock stars (purple) randomly drawn from different stellar population tracers -- uniform random (top left), 3.4 \mum near-IR tracing bulk stellar mass (top right), the 24 $\mu$m-corrected far-ultraviolet (FUV) tracing the $<100$ Myr star-formation (bottom left) and 24 $\mu$m-corrected H$\alpha$ tracing the $<10$ Myr star-formation (bottom right). Each map shows a population of $N=500$ randomly-drawn stars. The maps are themselves shown in grey-scale. Orange contours denote H$_2$ clouds with densities $>$4.3 \Msunpc from the ACA survey. The figure visualizes where stars from different progenitor populations occur, with their corresponding densities shown in Figure \ref{fig:mocks}.}
    \label{fig:mockmaps}
\end{figure*}
\begin{figure*}
    \includegraphics[width=\textwidth]{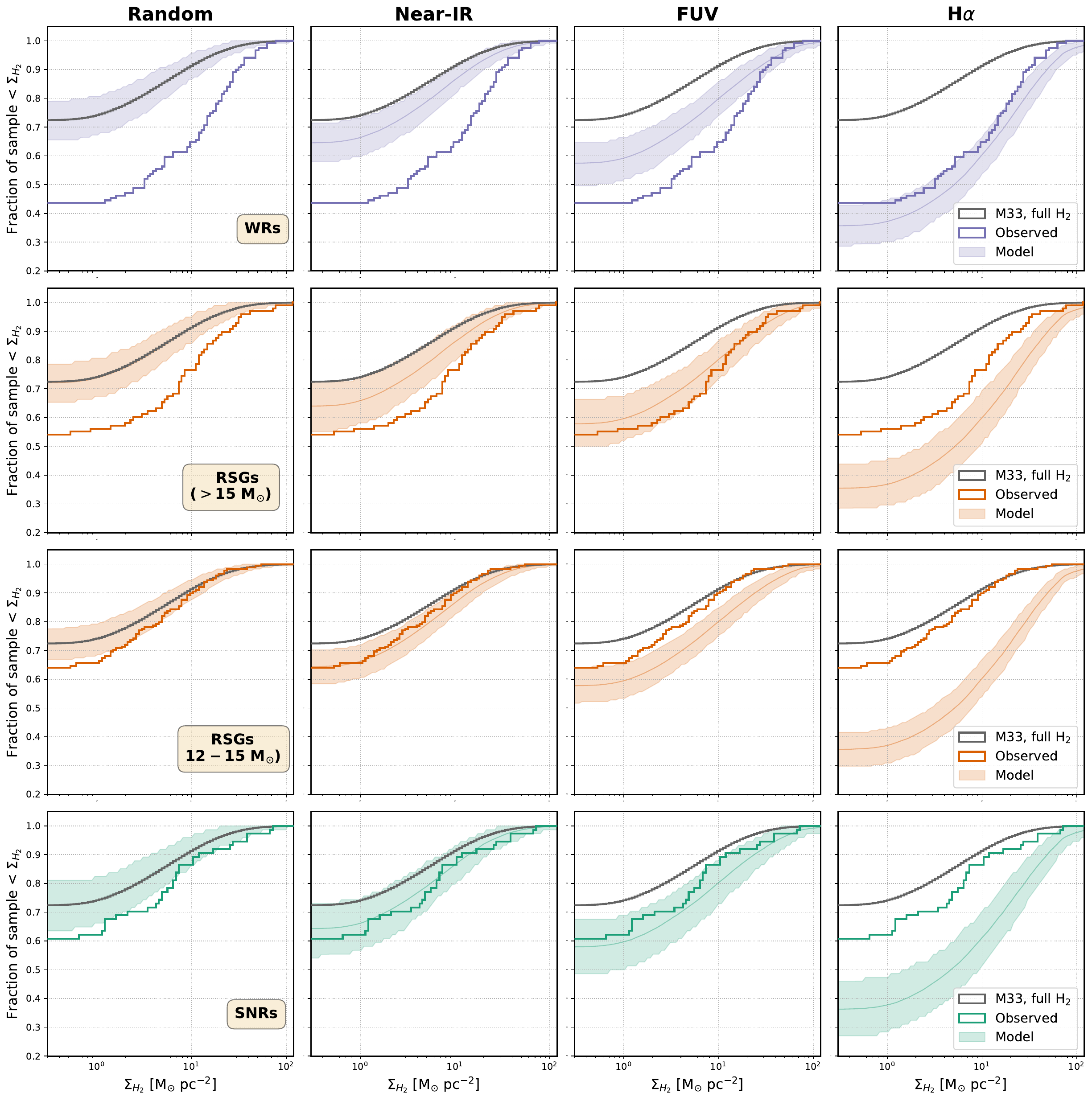}
    \caption{Cumulative density histograms of the observed and modeled populations of RSGs, WRs, and SNRs from Section \ref{sec:sfrtracers}. Solid grey \textbf{curves} in each panel denote the bulk H$_2$ cumulative distribution in M33 within the ACA area. Solid colored lines denote the H$_2$ surface density \textbf{distribution} of the star category. Colored shaded region shows the 5th-95th percentile region of the mock populations drawn according to Section \ref{sec:sfrtracers}. }
    \label{fig:mocks}
\end{figure*}
Are the above distributions therefore a consequence of the progenitor properties of WRs, RSGs and SNRs? We test this by creating mock populations of these stars from different star-formation tracer maps and comparing their density distributions with what is observed. The primary goal is not to determine the age distribution of RSGs, WRs and SNRs (which has been done elsewhere), but rather to determine the degree to which their observed ISM distribution is \emph{astrophysical} in nature, as opposed to a random alignment. This will also assess the stochasticity of the CDFs due to the finite sample sizes. 

To create the mock populations, we use maps of tracers of star-formation on different timecales. The probability ($p_i$) of a star occurring in pixel $i$ is weighted by the pixel luminosity in the tracer map (listed below) with respect to the total map luminosity, i.e., $p_i = L_i/\sum L_i$\footnote{Note that by definition we are drawing stars in a light-weighted manner, not mass-weighted, meaning the drawn populations are biased towards younger ages instead of sampling the underlying star-formation history, which requires more detailed modeling. Thus, in general, we are tracing progressively lower mass/older stellar populations as we go from H$\alpha$ to FUV to near-IR light, but the resulting stars have a light-weighted mass-range, which may not accurately reflect the specific mass range of our RSGs, WRs and SNRs.}. For each category, we generate populations from four different tracers as follows: 
\begin{enumerate}
\item \textbf{H$\alpha$+24$\mu$m} -- The H$\alpha$ emission line traces stellar populations younger than 10 Myr, with a mean age of 3 Myr \citep{Kennicutt2012}. We use \emph{MIPS} 24 $\mu$m maps \citep{Verley2007} to correct the continuum-subtracted H$\alpha$ map of M33 from \cite{HW2000} for extinction\footnote{The native maps in emission measure (EM, cm$^{-6}$ pc) units were converted to flux density ($S_{\mathrm{H}\alpha}$, ergs s$^{-1}$ cm$^{-2}$ arcsec$^{-2}$) units using the relation $EM = 4.858 \times 10^{17} T_4^{0.9} S_{\mathrm{H}\alpha}$ (\url{https://www.astronomy.ohio-state.edu/pogge.1/Ast871/Notes/Rayleighs.pdf)}, where we assume temperature $T_4 = T/10^4\mathrm{K}=1$.}. All maps are convolved and regridded to the ACA 12\arcsec resolution. We follow the extinction-correction prescriptions in \cite{Belfiore2023} to carry this out (see Appendix \ref{sec:extcorrfuvha} for details).
\item \textbf{FUV+24$\mu$m} -- Stars younger than 100 Myr account for about 90\% of UV emission, with a mean light-weighted age of about 10 Myr \citep{Kennicutt2012}. We use \emph{GALEX} FUV maps \citep{GALEXM33}; both convolved and regridded to the ACA 12\arcsec resolution, and extinction-corrected using MIPS 24 $\mu$m using the prescriptions in \cite{Belfiore2023}, as discussed in the Appendix \ref{sec:extcorrfuvha}.
\item \textbf{Near-IR} -- A population tracing the bulk stellar mass of the disk, which we can treat as a more realistic `random' population, i.e., stars follow the actual light profile of the galaxy. Analyses of resolved stellar populations in nearby galaxies have shown that near-IR bands are primarily dominated by red-giant and asymptotic-giant branch (AGB) stars in stellar populations with ages up to a few Gyrs \citep[$\gtrsim$2\Msun;][]{Dalcanton2012}, although some contribution from evolved massive stars such as RSGs, red core-He-burning and high-mass AGB stars are also expected \citep{Melbourne2012, Melbourne2013}. We use the bulk stellar mass map computed from the WISE W1 band (3.4 $\mu$m) of M33 to trace locations of stars for this case \citep{Leroy2019}. We refer the reader to Appendix \ref{appendix:volumedensity} for details of the conversion from W1 band map to stellar mass. We use the common astrometric and beam-matched 7.5\arcsec\ maps assembled in the z0MGs survey \citep{Leroy2019} and convolve to the 12\arcsec\ resolution and pixel scale of the ACA map. 
\item \textbf{Random} -- A completely random population, assuming each pixel has the same probability of hosting a star (i.e. $p_i = 1/\sum L_i$). This is the simplest control case -- the positions of stars in our map cannot be 100\% random, since they will follow the galactic substructure, but this gives a basic `control' sample for comparison to the other populations.

\end{enumerate}

The spatial distribution of the mock stars is shown in Figure \ref{fig:mockmaps}. Stars drawn from ``FUV'' and ``H$\alpha$'' maps (iii and iv) show more concentration along the star-forming and \hii\ regions, with a small fraction appearing in the inter-cloud and inter-arm regions. As expected, the concentration of stars in \hii\ regions is much higher when drawn from the H$\alpha$ maps than the FUV maps. In contrast, stars drawn from ``Random'' and ``near-IR'' are nearly uniformly distributed throughout the ACA region, with the near-IR population slightly more weighted towards the M33 center. However, even some of these random/near-IR stars fall within H$_2$ contours, which demonstrates how a fraction (seen by the limiting fractions in Figure \ref{fig:mocks}) of our samples could \emph{appear} to be associated with molecular clouds by chance alignment at our 50 pc scales even if they are not physically associated. 

For each tracer (i--iv) and star category (WRs, RSGs and SNRs), we generate 500 mock populations each of size $N$ (where $N$ is the observed sample size of the star category given in Table \ref{tab:densities}). We then compare the median, 5th and 95th percentile range of H$_2$ density of these randomly drawn stellar populations from each tracer to the observed RSG, WR and SNR distributions. Note that we only compare with H$_2$, since the differences between the \hi or \hi+H$_2$ CDFs of the three categories are less prominent than H$_2$ alone. Figure \ref{fig:mocks} shows the comparison of the H$_2$ CDFs between our randomly drawn mock populations and the observed stars. The shaded (2$\sigma$) indicates the variance due to repeat sampling, with larger variance seen for smaller samples. The CDFs of the mock populations behave similar to the observed ones, flattening to 1 at the maximum pixel values, and to a limiting value at pixel values  approaching the completeness limit of the CO map. 

We find in Figure \ref{fig:mocks} that the H$_2$ CDF of the WRs is most consistent with the mock CDFs drawn from the H$\alpha$ maps, and in tension with the other tracers. This is consistent with the results in Section \ref{sec:stats} and Table \ref{tab:stats}, which showed that H$_2$ CDF of the WRs is statistically different from the bulk M33 H$_2$ distribution. The mock CDFs also reproduce a fraction of stars with no detectable H$_2$, similar to the WRs. This indicates that H$_2$ distributions of WRs and the general $<$10 Myr stellar population in M33 are statistically similar, with a substantial fraction not coincident with molecular gas. The mock CDFs in Figure \ref{fig:mocks} do indicate that stars drawn from H$\alpha$+24$\mu$m are associated with slightly denser H$_2$ than the WRs. In fact, as shown in Appendix \ref{sec:extcorrfuvha} and Figure \ref{fig:fuvha}, CDFs of stars drawn from H$\alpha$ without 24$\mu$m correction are more consistent with the WR CDF, although both these cases are consistent with WRs within the shaded error regions. We return to this point in Section \ref{sec:discussion}.

In contrast to WRs, the H$_2$ CDFs of the RSGs are more consistent with the FUV and near-IR-drawn populations. Specifically, the $>$15 \Msun\ RSGs are best-reproduced by the FUV light, while 12--15 \Msun\ RSGs are reproduced partly by the near-IR light and partly by purely random sampling. This is consistent with the fact that FUV emission is weighted by younger stars than those contributing to 3.4$\mu$m emission, and further supports that lower mass stars are less likely to be associated with H$_2$. Interestingly, the 12--15 \Msun\ bracket is more consistent with random/near-IR than the younger (FUV/H$\alpha$) tracers. We expect 12--15 \Msun\ stars to have ages of $\sim$10-20 Myrs \citep{Ekstrom2012}, while the dominant contribution at $\lambda >1$ $\mu$m is from old stellar populations with light-weighted ages approaching a few Gyrs \citep[assuming star-formation histories typical of local universe galaxies, see][]{Conroy2013}. This likely implies that by the time  $<$15 \Msun\ stars are evolving well beyond the main sequence, they are almost unassociated with H$_2$ gas, and any spatial correlation with H$_2$ is statistically indistinguishable from that of the bulk older stellar population. The tension between the $>$15 \Msun\ RSGs and H$\alpha$ in contrast with the WRs is likely due to WRs having higher progenitor masses than RSGs. Stellar evolution models predict that at higher masses (and metallicities), the effects of binarity, rotation and mass-loss limits the production of cool RSGs with respect to WRs \citep{BPASS, Massey2021b}.

Overall, the analysis in this section drives home two important points -- that the density distributions of our stars in Figure \ref{fig:3paneldensities} 
 and \ref{fig:3paneldensities_rsgs} are indeed statistically unique  and not just due to chance alignment with the ISM (consistent with Section \ref{sec:stats}), and that their observed density distributions are \emph{astrophysical} in nature, consistent with the densities in which their general progenitor populations would also be evolving, with some stochasticity due to finite sample sizes. 

\subsubsection{Density distribution around the SNRs} \label{subsec:snrs}
The analysis in the above sections provides some context for understanding the \hi and H$_2$ CDFs of SNRs. We see from Figures \ref{fig:3paneldensities}--\ref{fig:mocks} that the \hi and H$_2$ CDFs of SNRs are more similar to the lower mass RSGs (9-15 \Msun) than the WRs or high-mass RSGs. Notably, the SNRs are inconsistent with being drawn purely from the H$\alpha$-emitting population. 

These results are overall consistent with the primarily low progenitor masses of SNRs, indicated by the surrounding star-formation histories\footnote{Note that `progenitor masses' derived this way are a statistical translation of the age distribution of stars around the SNRs into a single stellar-mass. They are not a direct measurement of the progenitor mass, which is difficult to obtain except for young ejecta-dominated SNRs, or ones with light echoes. The masses can have uncertainties exceeding a factor of 2 depending on the stellar age distribution. However, this still remains the \emph{only} way to obtain reasonable estimates of progenitor masses of individual extragalactic SNRs.}\citep{Jennings2014, DiazRodriguez2018, Koplitz2023}. The progenitor mass distribution peaks at $\sim$10 \Msun, with a small tail towards 30--40 \Msun. According to \cite{Koplitz2023}, about 40\% of the SNRs having masses $<$8 \Msun, which may indicate SNe Ia or delayed core-collapse SN origins \citep{Zapartas2017}. We do find in Appendix \ref{sec:snria} and Figure \ref{fig:snrsia} that a 40\% Type Ia contribution (assuming they are drawn from the near-IR map) produces a slightly better match to the observed H$_2$ CDF of SNRs in Figure \ref{fig:mocks}. In contrast, all the WRs come from high-mass ($>$30 \Msun) progenitors, so they are more likely than the general SNR population to be near dense H$_2$ gas. 

A secondary effect could be that some of the SNRs have cleared out H$_2$ within their current shock radius, though this is difficult to confirm at our 50 pc resolution, which is larger than the mean diameters of M33 SNRs \citep{Lee2014, White2019}. As mentioned in Section \ref{sec:cavities}, we have ongoing $\sim$pc-scale ALMA  observations of M33, which we will use in future papers to better confirm the fate of the gas within the SNR volume. 

\section{Where do stars explode in the ISM? -- The general picture} \label{sec:discussion}
In this section, we weave the above results into a general picture of where stars are exploding and the physical mechanisms driving these trends, the implications for subgrid models of feedback in hydrodynamical simulations, and comparison with existing literature on SNe and SNR environments. 

\subsection{Correlation between ISM density and progenitor age} \label{sec:densityage}
\begin{figure}
    \centering
    \includegraphics[width=0.94\columnwidth]{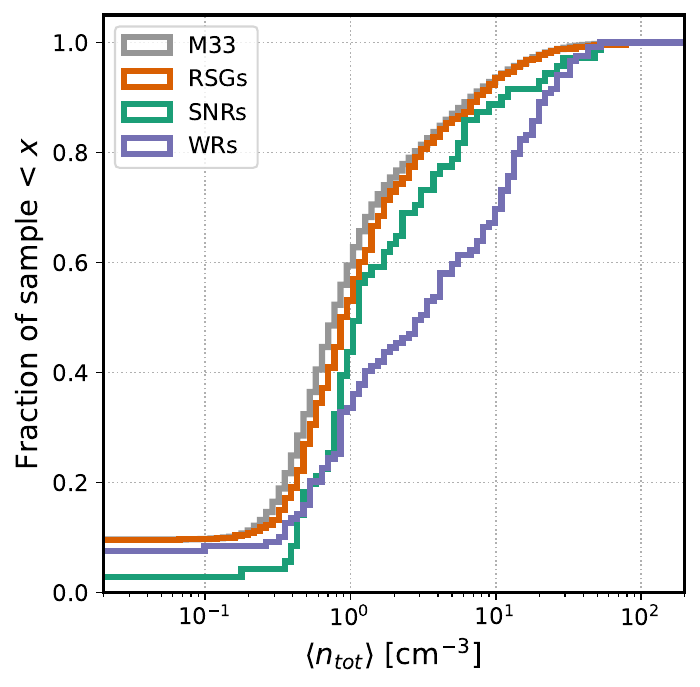}
        \caption{CDFs of the average cold (\hi + H$_2$) gas volume densities for WRs, SNRs and RSGs, derived according to Appendix \ref{appendix:volumedensity}. Grey CDF denotes the bulk cold density distribution in M33. See Section \ref{sec:densityage} for details.}
            \label{fig:ntot_stars}
\end{figure}

\begin{figure}
    \centering
    \includegraphics[width=\columnwidth]{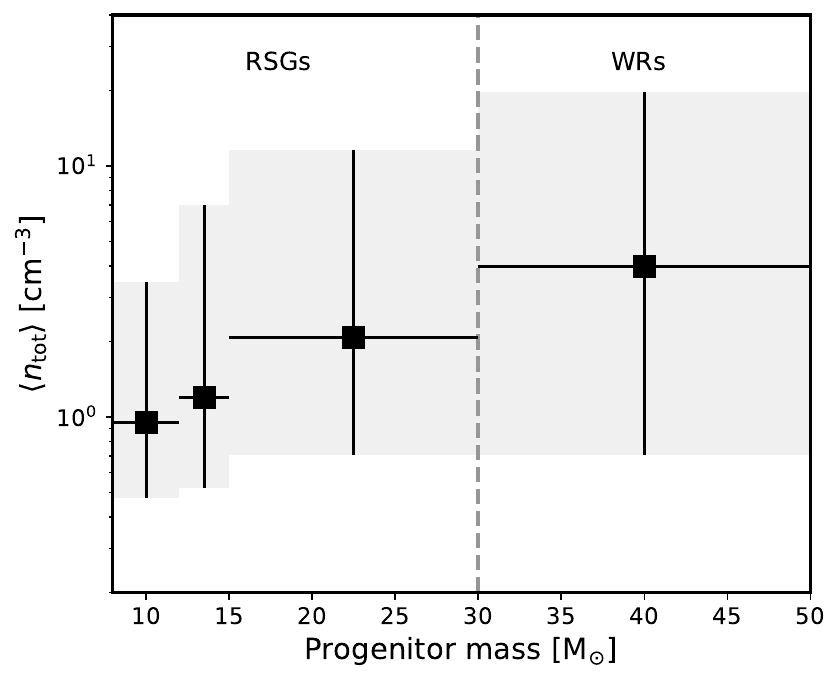}
    \caption{Average cold (HI + H$_2$) gas volume density, sampled at 48 pc resolution, vs progenitor mass range of red supergiants and Wolf-Rayet stars. The grey shading indicates the 16th-84th percentile range. This gives a rough idea of the progressively increasing density expected around massive stars with younger ages.}
    \label{fig:ntot_progmass}
\end{figure}
One of the major implications of our results is that there is a clear progenitor age/mass dependence on the ISM densities where stars explode.\ WRs, which are the most massive (and shortest-lived) of the three categories, have the highest sample fraction (55\%) evolving in dense, molecular gas-rich environments. More luminous (and thus more massive and shorter-lived) RSGs also have a higher fraction (44\%) of stars evolving in dense, molecular-gas dominated environments than less luminous RSGs. The progenitor mass distribution of SNRs peaks at 10 \Msun (Section \ref{sec:sfrtracers}) which is much lower than WRs, and there may be some Type Ia SNRs with lower mass progenitors. We therefore see a larger sample fraction of SNRs with non-detections than WRs (Figures \ref{fig:3paneldensities}, \ref{fig:3paneldensities_rsgs}). Similar trends are also observed for these stars in relation to age-sensitive stellar population tracers as discussed in Section \ref{sec:sfrtracers}. 

The simplest physical interpretation of this trend is that more massive stars have shorter lifetimes, and are thus more likely to explode close to their birth clouds, or before the clouds have been dissipated. We caution however that this fraction of stars correlated with dense gas at our 50 pc resolution should be treated as an upper limit to the actual number of stars that will interact with dense gas. While the H$_2$ cloud densities on these spatial scales may not be significanty modified by turbulence before our WRs and RSGs explode (Section \ref{sec:obs-why}), they may get modified in regions of high gas density and dense stellar clustering (e.g. NGC 604) by powerful pre-SN feedback and frequent SN episodes. Thus some of the massive progenitors in such dense regions may explode in evacuated regions (which can be better examined with higher-resolution CO data as discussed in Section \ref{sec:cavities}), while others might still end up interacting with dense gas (indeed many SNRs have been observed to be interacting with dense molecular gas, Section \ref{sec:compareSNstudies}).

We can also express the surface density distributions in units of volume densities to make them more accessible to simulations. In contrast to surface densities however, which are directly measured from the line fluxes, volume densities require some assumption about the line of sight distribution of gas, and thus are somewhat model-dependent. Any line of sight through the turbulent ISM disk of a star-forming galaxy will coincide with bubbles and inhomogeneities that are not captured in projection, but simulations have shown such disks evolve to an equilibrium state where the gravitational pressure of dark matter, stars, and gas is balanced by thermal and turbulent pressure primarily from stellar feedback \citep{OML10, OS11, Kim2011}. Assuming such an equilibrium in M33, we derive \hi and H$_2$ scale heights and volume densities in our ACA region in Appendix \ref{appendix:volumedensity} in each pixel given the locally measured stellar mass, gas surface densities, and velocity dispersions. These should be treated as densities averaged over 50-pc beams.

Figure \ref{fig:ntot_stars} shows the volume densities of three star categories. All stars have average ambient densities between $0.1-50$ cm$^{-3}$. The general trends from Figures \ref{fig:3paneldensities} and \ref{fig:3paneldensities_rsgs} are still noticeable. WRs once again evolve in the highest range of densities compared to the SNRs and RSGs. In Figure \ref{fig:ntot_progmass}, we show the average densities for WRs and RSGs, specifically the median and the 16th-84th percentile range vs their approximate range of single stellar progenitor masses. We also subdivide the RSG sample into the three mass bins of $8-12$ \Msun, $12-15$ \Msun and $>15$\Msun\ as described in Section \ref{sec:physcorr}. While each category evolves in an order-of-magnitude range of densities, one can clearly see a systematic increase in average densities around the stars with higher progenitor mass.

\subsection{Significant fraction of future SNe outside molecular clouds} \label{sec:lowdensitySN}
\subsubsection{General statistics} \label{sec:genstats}
Another key implication from Figures \ref{fig:3paneldensities}, \ref{fig:3paneldensities_rsgs} and Section \ref{sec:physcorr} is that a significant fraction of future core-collapse SNe will occur in low-density inter-cloud regions, dominated by atomic gas. Only about 30-40\% of the RSGs are in pixels that contain H$_2$, and the rest in \hi-dominated pixels. A similar fraction of low-density environments are also inferred for SNRs, which in M33 are somewhat dominated by lower-mass ($\sim$10 \Msun) progenitors \citep{DiazRodriguez2018,Koplitz2023}. The results are consistent with MC22, who also showed that almost 50\% Type II SNe (primarily arising from red and yellow supergiant stars) were not coincident with detectable molecular clouds. 

We note however that the shocks from future SN explosions of these stars in the intercloud region can still travel tens of pc and interact with nearby H$_2$ gas clouds. This association can be seen somewhat by eye in Figure \ref{fig:maps}, where many SNRs are not coincident with, but are near a molecular cloud. In terms of feedback, the ability of SNe to disperse dense gas reduces with distance from the molecular cloud \citep{Hennebelle2014, Iffrig2015}, although the edge interaction can still accelerate cosmic rays \citep{Sano2021} and/or compress gas potentially to star-forming densities \citep{Cosentino2022}, but with a lower angular filling fraction than a fully embedded spherical shock.

\begin{figure}
    \includegraphics[width=\columnwidth]{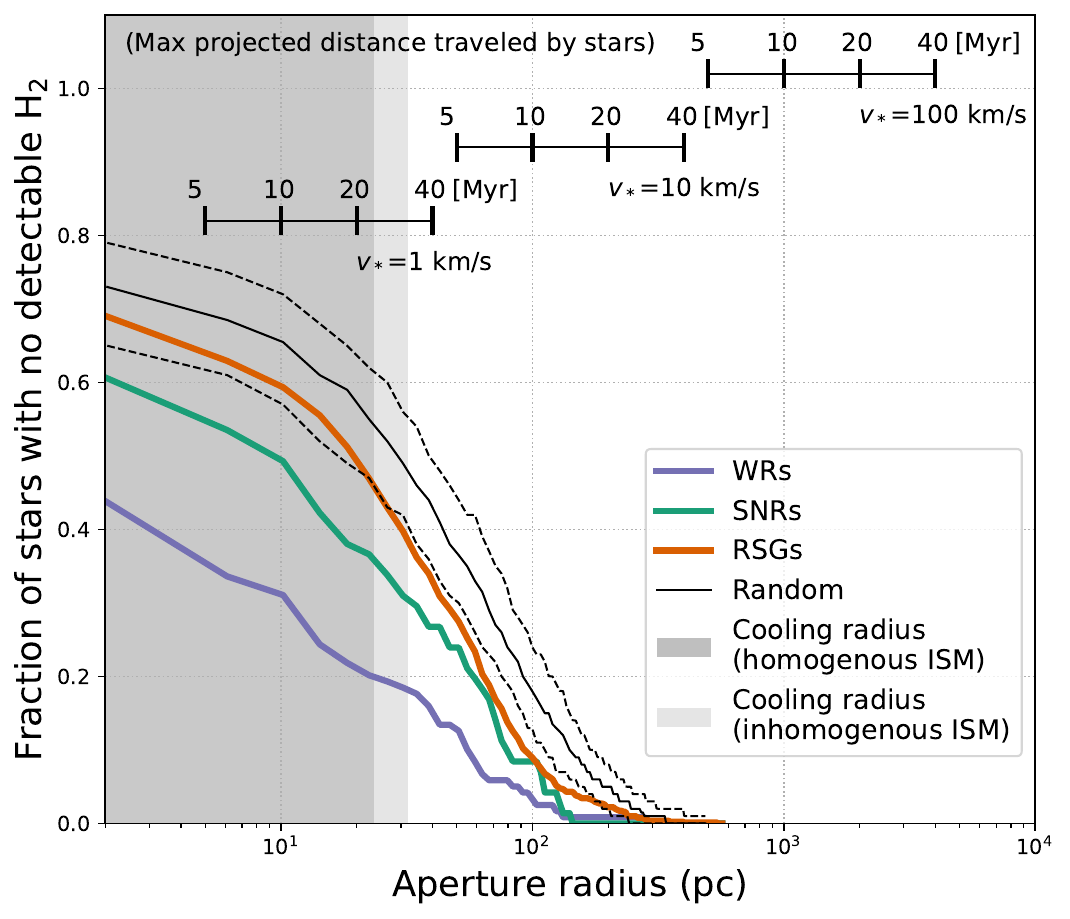}
    \caption{Fraction of stars in the samples of WRs, RSGs and SNRs without any detectable pixels of H$_2$ (traced by CO 2-1) within an aperture of given search radius. For comparison with stars uncorrelated with molecular gas, we also show the non-detection fraction of a mock population drawn randomly in the ACA area (the ``Random'' case from Figures \ref{fig:mockmaps} and \ref{fig:mocks} ) with thin solid and dashed lines, with solid showing the median, and dashed showing the 5th and 95 percentile bounds. The shaded region denotes the cooling radii for a supernova shock evolving in density = 1 \cmc and metallicity = 0.6 $Z_\odot$ for a uniform ISM (dark shaded region) and inhomogeneous ISM (light shaded region; Eq. \ref{eq:Rc}).
    The horizontal scales show the maximum projected distance that a star would travel within its lifetime (vertical ticks) for velocities = 1, 10 and 100 km/s. Note that the vertical positioning of the scales are arbitrary.}
    \label{fig:aperture}
\end{figure}

We therefore assess the H$_2$ non-detection fraction around stars as a function of a search radius (called `aperture radius', following the nomenclature of MC22) in Figure \ref{fig:aperture}. Specifically, we plot the sample fraction containing no detectable H$_2$ pixels within circles of increasing radii from the central star. To distinguish from the H$_2$ distribution of stars uncorrelated with molecular clouds, we also show in Figure \ref{fig:aperture} the aperture non-detection fraction for the ``Random'' population introduced in Section \ref{sec:sfrtracers}.

Figure \ref{fig:aperture} highlights the importance of spatial resolution for characterizing the physical association of SNe with molecular gas (a point we will return to in Section \ref{sec:cavities}). The sample fractions of non-detection are the highest for each category at the ACA map pixel scale ($\sim$4 pc), and then slowly decreases as we increase the aperture radius. At $\gtrsim$300 pc, every star has some detectable H$_2$ pixel. Figure \ref{fig:aperture} also reaffirms that the distribution of non-detection fraction vs aperture radius is different from the case of stars randomly aligned with respect to molecular clouds. Similarly, MC22 showed in their Figure 2 and Table 2 that their H$_2$ detection fraction around core-collapse SNe increases from 60\% at 150 pc resolution to 94\% at 1 kpc resolution. Coincidentally, this result bears similarity to the findings that CO and star formation are strongly spatially correlated  on large scales, but becomes de-correlated on the scales of individual giant molecular clouds \citep[e.g.][]{Schruba2010, Kruijssen2019}.

We then compare the sample fractions of non-detection vs radii in Figure \ref{fig:aperture} with the typical cooling radius expected from a 10$^{51}$ erg SN explosion in atomic gas (n = 1 cm$^{-3}$). The SNR cooling radius is an important resolution criterion in ISM simulations to identify whether SN feedback is implemented primarily in the form of thermal energy or momentum \citep{Hopkins2013, Martizzi2015, Kim2015momentum}. By the time the SNR has expanded to its cooling radius, which typically marks the end of the adiabatic Sedov-Taylor stage, it has deposited $\gtrsim$50\% of its final momentum into the surrounding gas, and has begun quickly radiating away thermal energy. 

\cite{Kim2015momentum} expressed the cooling radius ($R_c$; or the shell formation radius) fitted to numerical simulations of SNRs from 10$^{51}$ erg SN explosions as 
\begin{equation}
    R_c = 
    \begin{cases}
    (22\ \mathrm{pc})(Z^2+0.04^2)^{-0.061}n_{1}^{-0.43}\\ 
    \qquad \qquad \qquad \qquad \qquad \quad \text{(homogenous ISM)} \\[0.3cm]
    (30\ \mathrm{pc})(Z^2+0.04^2)^{-0.061}n_{1}^{-0.46}\\
        \qquad \qquad \qquad \qquad \qquad \quad (\text{inhomogenous ISM}) \\
    \end{cases}
    \label{eq:Rc}
\end{equation}
where $Z$ is the ISM metallicity in units of $Z_{\odot}$ and $n_{1} = n/1\mathrm{cm}^{-3}$ is the average gas number density (which is characteristic of the median densities inferred from Figure \ref{fig:ntot_stars}). The metallicity dependence is taken from \cite{JGKim2023}.
The cooling radius is larger in an inhomogeneous ISM as the SNR forward shock escapes more quickly through the larger volume-filling low-density channels. We assume a typical atomic gas density of 1 \cmc at Z = 0.6 Z${\odot}$, which gives $R_c = 23.4$ pc (homogenous) and $R_c = 31.9$ pc (inhomogeneous), shown as grey shaded regions in Figure \ref{fig:aperture}. 

We find that 10-20\% WRs, 21-34\% SNRs, and 30-50\% RSGs have no detectable H$_2$ within the typical cooling radii shown in Figure \ref{fig:aperture}. In other words, the blastwaves from the explosions of these stars in the inter-cloud media, assuming typical ISM densities of $\sim$1 \cmc, are unlikely to affect nearby molecular gas before the onset of significant thermal energy losses in the shock. They may of course still interact if the inter-cloud densities are much smaller (e.g. $R_c \approx 63-170$ pc for the homogeneous case with $n_1=0.1-0.01$ at $Z=0.6$ Z$_{\odot}$).

\subsubsection{Long lifetimes, runaways, cloud dispersal?} \label{sec:reasonforlowdensities}
While the H$_2$ non-detection fraction of RSGs and SNR progenitors could be explained by them simply drifting away from their birth clouds over their long lifetimes (upto $\sim$40 Myrs), it is intriguing why a substantial fraction of WRs have no detectable H$_2$ despite their much shorter single stellar ages of a few Myrs \citep{BPASS, Dorn2018}. Since we do not have detailed 3D kinematic information for the WRs \citep[see][for a discussion of the complexities of radial velocity measurements of WRs]{Neugent2014}, we make use of the separation information in Figure \ref{fig:aperture} and some published observations to assess possible scenarios.

One possibility is that these WRs had escaped their birth clouds at high velocities acquired from either dynamical interactions or supernovae (believed to be the cause of runaway OB stars; \citealt{Fujii2011}). We assume the separation between the WR and natal cloud is $d = v_* t_{\mathrm{wr}}$, where $v_*$ is the relative velocity between the WR and the cloud, and $t_{\mathrm{wr}}$ is the age of the WR progenitor. In a single stellar population at LMC-like metallicity, WRs have an average age of $t_{\mathrm{wr}}=4$ Myrs, while in a 100\% binary population, $t_{\mathrm{wr}}=10$ Myrs, with the majority being within 16 Myrs \citep{BPASS, Dorn2018}.  Therefore, the 20\% of WRs in Figure \ref{fig:aperture} that have no detectable H$_2$ within atleast 32 pc (the assumed cooling radius in the homogenous case in Section \ref{sec:genstats}) would need $v_* > 8$ km/s if they were single stars, and $v_*>3.2$ km/s if they were in an interacting binary system\footnote{We use `$>$' because the actual velocity needed to produce the projected separation may be larger if the velocity vector is not in the plane of sky; in addition; the WR may not have originated in the nearest H$_2$-detected pixel, but farther out.}. The 5\% WRs in Figure \ref{fig:aperture} that are even more remote (with no detectable H$_2$ within $>$100 pc) would require $v_* > 25$ km/s (single) and $v_* > 10$ km/s (binary). These velocities are not unlike those predicted by models of walkaway and runaway stars produced by dynamical ejections and supernova \citep[e.g.,][]{Oh2016, Renzo2019, DorigoJones2020}. Such runaways have been considered in simulations as a way to boost outflows in low-mass galaxies \citep[e.g.][]{Andersson2020, Steinwandel2020}. A non-trivial number of runaway OB stars are known to exist in the Galaxy and Magellanic Clouds, though the estimated fraction of the total population varies in the literature \citep[e.g.][]{Blaauw61, deWit2005, Gvaramadze2012, Lamb2016}. Therefore it is not implausible that the H$_2$ non-detection WRs are a result of them being displaced from their birth locations at some walkaway/runaway-star-like velocity.

It is also possible that the parent clouds of WRs were quickly destroyed by pre-SN and prior-SN feedback. Molecular clouds have been estimated to disperse on 1-5 Myr timescales after emergence of massive stars, based both on observations of giant molecular clouds and H$\alpha$-emitting stellar populations of nearby galaxies \citep[e.g.][]{Chevance2020} as well as detailed high-resolution simulations of turbulent molecular clouds \citep[e.g.][]{Grudic2022}.\ Nearly 79\% of the WRs with H$_2$ non-detection (i.e. 42 out of 53 WRs) are coincident with known OB associations according to \cite{Neugent2011}, who cross-matched their WR sample with the OB associations of \cite{Humphreys1980}. The H$_2$ detected WR population also has a similar ($\sim$80\%) fraction of WRs coincident with OB associations (50 out of 63 WRs). So it is possible that the WR population with H$_2$ non-detections are still within their parent OB association, and since the typical stellar velocity dispersions in OB associations is $\sim$4 km/s \citep{Melnik2017,Wright2022}, these WRs may not have acquired the velocities estimated in the previous paragraph to produce the observed cloud-star separations in Figure \ref{fig:aperture}, which leaves the cloud destruction scenario by pre-SN/prior-SN feedback a more plausible explanation. The remaining 21\% of WRs with neither detectable H$_2$ nor identified within any known OB associations may be stronger candidates for walkaways/runaways. 
\begin{figure*}
    \includegraphics[width=\textwidth]{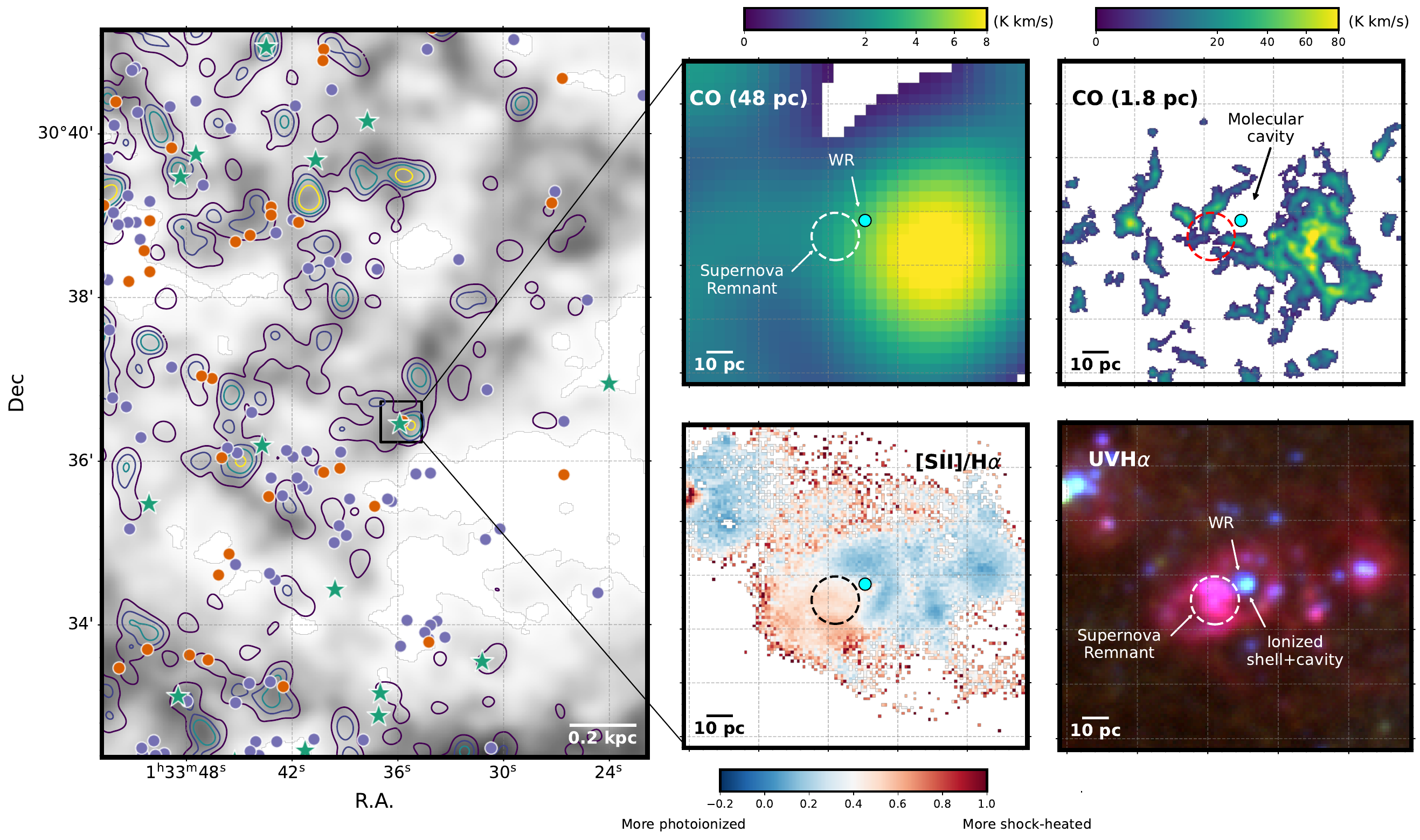}
    \caption{Zoom-in region of the Wolf-Rayet/SNR pair from Figure \ref{fig:maps}, demonstrating the benefit of CO observations with increasing spatial resolution. (\textit{Left}) Small patch of the ACA CO(2-1) + 21 cm map at 12\arcsec resolution from Figure \ref{fig:maps}, more centered on the WR/SNR pair. Symbols denoting RSGs (orange), SNRs (green) and WRs (purple) are the same as Figure \ref{fig:maps}. The black box is the $\sim$(122 pc)$^2$ region detailed in the smaller panels. (\textit{Top middle}) Same low-resolution (12\arcsec $\approx$48 pc) CO (2-1) map as the left panel, but zoomed in. The WR and SNR are denoted. (\textit{Top right}) High-resolution  (0.45\arcsec $\approx$1.8 pc) ALMA CO  map of the same region (Section \ref{sec:highresalma}), showing more detailed substructure of the molecular ISM. The WR is clearly located in a vacant cavity in the ISM. (\textit{Bottom middle}) Continuum-subtracted [\ion{S}{2}]/H$\alpha$ from narrowband images taken by Massey et al 2007. The colorscale is split at a value of 0.4, commonly used to distinguish photoionized (bluer) from shock-heated (redder) regions. We clearly see the SNR region as a shock-heated nebula, with the WR embedded in a more ionization-dominated region. (\textit{Bottom Right}) RGB image showing broadband $U$ (blue), V (green) and narrowband H$\alpha$ (red) from Massey et al (2006) and Massey et al (2007). The Wolf-Rayet star is clearly visible as a hot star, and the SNR nebula is visible below it.  The higher-resolution CO map reveals the detailed substructure around the WR, particularly its evolution inside a cavity presumably driven by powerful pre-SN feedback.}
    \label{fig:zoomin}
\end{figure*}
\subsection{Cavities and substructures on scales $\ll$ 50 pc} \label{sec:cavities}
Although we are finding younger massive progenitors closer to dense gas, a caveat is the spatial resolution of 50 pc.  While this is higher than extragalactic SN studies (e.g., MC22), the 50 pc beam still only captures a relatively large-scale mean gaseous environment, and misses substantial structure and inhomogeneity on smaller scales. 

As a case study, we show in Figure \ref{fig:zoomin} a region of M33 (the black box from Figure \ref{fig:maps}) containing a giant molecular cloud, a WR \citep[J013335.73+30.3629.1 in][]{Neugent2011}  and an SNR \citep[L10-045 in][]{White2019}, where we also have pc-scale CO (2-1) data from the ALMA archive (Section \ref{sec:highresalma}).  The low-resolution ACA image shows the WR evolving in detectable H$_2$ and close to the peak H$_2$ emission of the cloud, but the higher resolution image shows the WR inside a $\sim$10 pc radius cavity. The peak emission comes from a neighboring cloud complex likely produced by ongoing cloud collisions \citep{Sano2021}. We also see a similar cavity in the H$\alpha$ image (bottom right panel), where the WR star is surrounded by an H$\alpha$ shell of similar extent as the H$_2$. The H$\alpha$ shell also shows [\ion{S}{2}]/H$\alpha$ $<0.4$, indicating it is photoionization-dominated as opposed to shock-heated \citep{Matonick1997}. This proves that it is a real region of gas deficit, most likely encased in a photoionized shell sweeping up surrounding CO-emitting H$_2$. Similar structures have been observed around WRs in the Galaxy \citep[e.g.,][]{Baug2019} and Magellanic Clouds \citep[e.g][]{Tarantino24}.

A $\sim$10 pc-radius cavity could have been produced by a previous SN explosion or pre-SN feedback. In the first of two possibilities, a 10$^{51}$ erg SN explosion will sweep up ambient gas until the shock velocity is reduced to the turbulent velocity dispersion ($\sigma$) of the ISM. Based on Eq 20 in \cite{Martizzi2015}, the maximum radius of the swept-up momentum-driven SNR shell is,
\begin{equation} \label{eq:Rsnr}
    R_{\mathrm{SNR}} = (12.5\ \mathrm{pc})\ n_{2}^{-0.39} \sigma_{5}^{-0.33}
\end{equation} 
for an ambient gas with density $n_{2} = n_0/10^2$ cm$^{-3}$, and turbulent velocity dispersion $\sigma_5 = \sigma/5$ km s$^{-1}$, where we normalized to 5 km/s which is the average value inside the inner 5 kpc of M33 \citep{Koch2019}. Eq \eqref{eq:Rsnr} shows that a SN explosion with typical energy output can easily produce a 10 pc-radius cavity inside molecular gas with typically observed densities and turbulence.  

Pre-SN feedback from the WR and its preceding O-star phase could also have also carved out the present-day cavity. The WR was not identified as part of any known OB association by \cite{Neugent2011}, and no catalogued star-cluster is in the cavity \citep{Sarajedini2007, Fan2014, Johnson2022}, so it is likely the WR star would be the primary power source\footnote{We also note that no other spectroscopically-confirmed O/B star from \cite{Massey2016} is in the vicinity, though completeness of the O-star catalog is still an issue.}. The expansion of a D-type ionization front around OB stars in a uniform ambient medium with density $n_0$ can be expressed as in \cite{Hokosawa2006} Eq (36) as
\begin{equation}
R_{\mathrm{II}}(t) = R_{\mathrm{st}}\left(1 + \frac{7}{2\sqrt{3}} \frac{c_{s}t}{R_{\mathrm{st}}}\right)^{4/7}
\label{eq:RII}
\end{equation}
where $R_{\mathrm{st}}$ is the Stromgern sphere given by 
\begin{equation}
R_{\mathrm{st}} = \left(\frac{3 Q_0}{4 \pi \alpha_B n_0^2}\right)^{1/3}
\end{equation}
where $\alpha_B = 3.023 \times 10^{-13}$ cm$^{3}$ s$^{-1}$ is the case-B recombination coefficient, $Q_0$ is the ionizing luminosity (s$^{-1}$), $c_s \approx (11.6\ \mathrm{km/s})(T/10^4\mathrm{K})^{0.5}$is the sound speed inside the \hii region. Based on Eq \eqref{eq:RII}, stars $>$30 \Msun with typical $Q_0=10^{49}-10^{50}$ s$^{-1}$ \citep{Sternberg2003} can easily produce ionized bubbles $>$10 pc inside gas with densities $n_0 = 10^2$ \cmc within $t\approx 0.4-1$ Myrs. These numbers are also consistent with results from more detailed numerical simulations \citep{JGKim2018, JGKim2021, JGKim2023}.

It is also possible that the cavity was produced by strong stellar winds, given the typical mass-loss rates in O-stars being in the range of $10^{-7}-10^{-5}$ \Msun yr$^{-1}$ at velocities $>$10$^{3}$ km s$^{-1}$, and around $10^{-5}-10^{-4}$ \Msun yr$^{-1}$ for WRs \citep{Smith2014}. \cite{Lancaster2021a} and \cite{Lancaster2021b} derived the radius of such wind-driven bubbles with the inclusion of cooling at the wind-shell interface as,

\begin{equation}
R_{b,\mathrm{mom}} = (3.55\ \mathrm{pc})\ \dot{M}_{w,-6}^{1/4} v_{w,3}^{1/4} n_{2}^{-1/4} t_6^{1/2} \\ 
\end{equation}

where $\dot{M}_{w,-6}=\dot{M}_w/10^{-6}$ \Msun yr$^{-1}$ is the mass-loss rate, $v_{w,3} = v_w/10^3$ km s$^{-1}$ is the wind speed, and $t_6 = t/10^{6}$ yr is the age of the bubble. A massive star with a wind luminosity  $\dot{M}_{w,-6}>10$ and $v_{w,3}>4$ can easily produce cavity with the observed extent, assuming same cloud density and driving age.

This cavity around the WR, which could have easily been produced by either pre-SN or prior-SN feedback, will have a significant impact on its future SN explosion. In the absence of a cavity, the thermal energy of the WR SN blast wave would have been efficiently radiated away on 6--12 pc scales (Eq \ref{eq:Rc}), limiting the spatial extent of its impact. In the case of explosion inside a cavity, where we assume the gas density is now quite low ($\lesssim$10$^{-2}$ cm$^{-3}$), the cooling radius is about 150 pc, which means the blast wave will easily retain its total energy until it impacts the surrounding cloud material. Interaction of the blastwave with the surrounding swept-up ISM could also trigger massive star-formation in the future, similar to the WR itself\citep[e.g.][]{Cosentino2022}.

The analysis in Figure \ref{fig:zoomin} further highlights the need for mapping the cold, dense ISM in these galaxies at pc-scale resolution in order to unveil the true non-detection fraction of dense gas around future SNe, and the distribution of high and low-density channels of ambient gas around massive stars through which their future SN blast-waves will eventually evolve. Such small-scale substructures could also be pervading other peaks of gas density seen in Figure \ref{fig:maps} where massive WRs and RSGs are coincident, but smeared out by our 12\arcsec resolution.

\subsection{Comparison of observed densities around SNe with simulations} \label{sec:disc:sims}
\begin{figure*}
    \centering
    \includegraphics[width=0.9\textwidth]{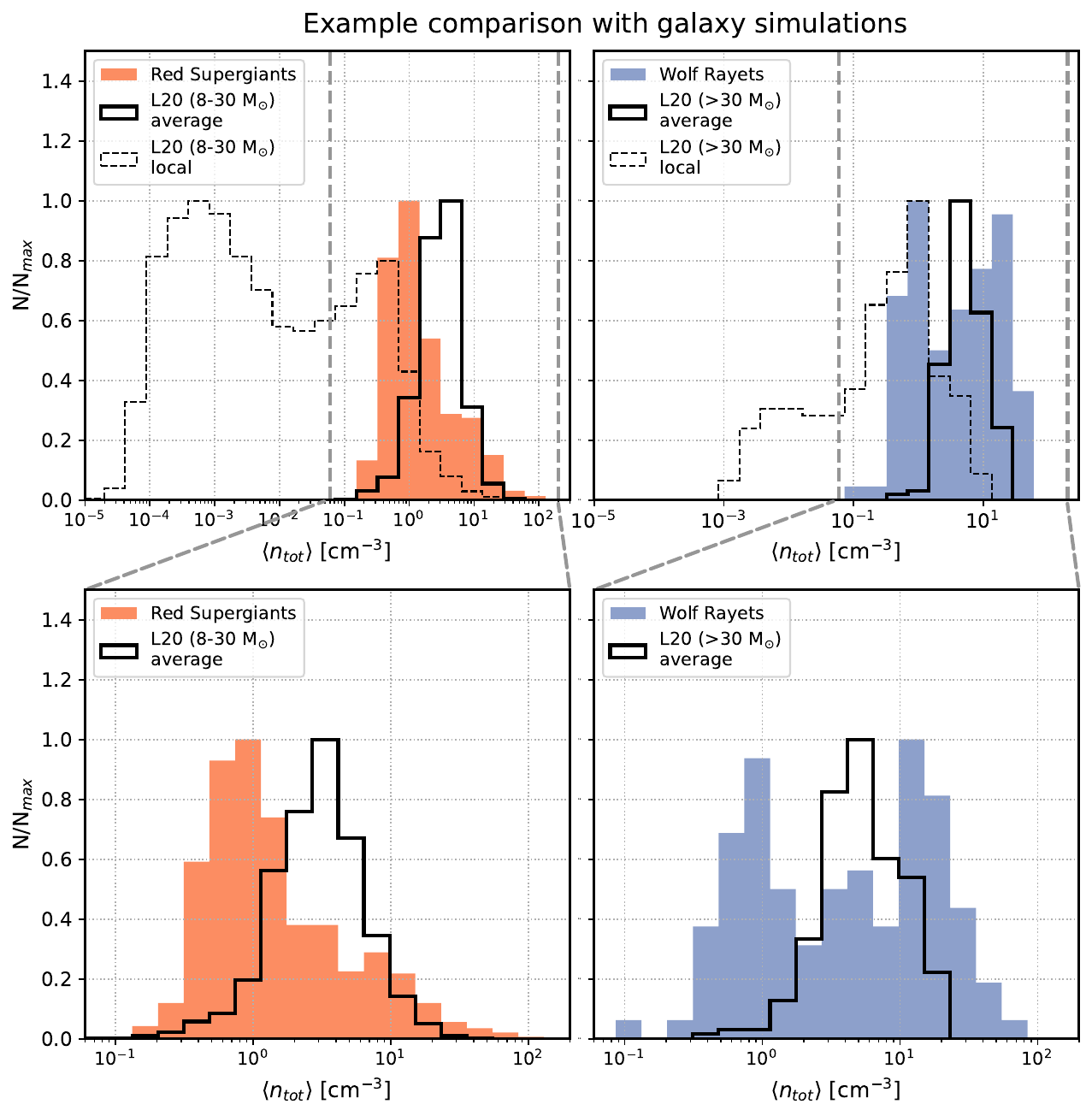}
        \caption{Example comparison of the \hi + H$_2$ gas volume density ($n_{tot}$) of RSGs (\emph{left} panels) and WRs (\emph{right}) at 50 pc resolution with the ambient SN densities from a high-resolution hydrodynamical simulation of a dwarf starburst galaxy \citep{Lahen2020}. Simulated densities were obtained for all SNe during the quiescent, pre-starburst phase between 70-140 Myrs \citep[See Fig 5 in][]{Lahen2020}, and divided into 8-30 \Msun and $>30$\Msun mass ranges to better compare with RSGs and WRs respectively. The observations, i.e. WRs and RSGs, are shown as colored histograms. The `local' theoretical densities (i.e. measured in the immediate region around the SN) are shown in the top panel as thin dashed histograms, while the `average' theoretical densities (measured from 2D-projected gas maps) are shown in thick histograms. Bottom panel zooms into the observed and the average density histograms, which are more directly comparable. The plot shows the feasibility of comparing ambient densities of simulations and observations as an independent way to assess the ISM and feedback physics (see Section \ref{sec:disc:sims} for a detailed discussion).}
            \label{fig:simvsobs}
\end{figure*}

In this section, we compare the observed densities of our WRs and RSGs from Figure \ref{fig:ntot_stars} (colored histograms) to the ambient SN densities from a high-resolution ($\sim$4 \Msun) simulation of a dwarf starburst galaxy from \cite{Lahen2020} mentioned in Section \ref{sec:data:sims}. The results of the comparison is shown in Figure \ref{fig:simvsobs}. Since our observed densities are derived from the projected ISM densities divided by a scale height (Section \ref{sec:densityage} and Appendix \ref{appendix:volumedensity}), we obtained average SN densities in a similar manner from the simulation, as described in Section \ref{sec:data:sims} (solid black histograms). For the sake of discussion, we show both the local (thin-dashed) and averaged (thick-solid) densities around simulated SNe in Figure \ref{fig:simvsobs}.

One will notice firstly that the local densities are much lower than the projected densities. This is a common prediction of high-resolution galaxy simulations where majority of stars explode in low densities carved by photoionization and SN clustering, and due to stellar motion away from the birth clouds \citep[e.g.][]{Hu2017, Lahen2020, Rathjen2021, Andersson2022}. Unfortunately, the immediate local densities are not measurable in observations where we always see the ISM and stars in projection. In addition, our ISM tracers of \hi and H$_2$ are mainly sensitive to gas densities $>0.1$ \cmc (since our main goal in this paper was to measure the quantity of dense gas around SNe). In future we can include recombination line and diffuse X-ray maps of galaxies to estimate the proportions of the ionized and hot ISM gas along the line of sight of stars.

Despite these limitations, Figure \ref{fig:simvsobs} shows how observations of the line of sight densities presented in this paper can serve as a possible quality assurance step in the SN physics of feedback models. For this particular simulation, we find that the range of average densities of the simulated and observed SNe are similar, with values around 0.1-100 \cmc. This is an encouraging sign that the highest-resolution simulations in the community, where feedback-driven ISM and star-formation is effectively generated from first-principles with little to no reliance on subgrid models, are converging with observations on similarly resolved scales.

However, the observations also reveal some subtle differences with the simulated densities. Most notably, on the right panels, we find that the bimodal WR density in the observations (with one peak at $\sim$ 1 \cmc associated with primarily atomic densities, and other at $\sim$ 10 \cmc associated with primarily molecular gas) is absent in the simulated ($>$30 \Msun ) stars, which instead show a unimodal distribution. On the other hand, on the left panels, we see that the distribution of the densities of the lower-mass stars (i.e. RSGs) have a similar unimodel shape as the 8-30 \Msun stars in the simulations. The peak of the RSG distribution however, appears to be lower by a factor of 3 compared to the simulation. Possible physical explanations for these discrepancies could be due to the strength of pre-SN feedback and degree of SN clustering in simulations \citep[e.g.][]{Gentry2017, Naab2017, Smith2018, Rathjen2021}, runaways \citep[e.g.][]{Andersson2020} and choice of numerical models \citep[e.g.][]{Hu2022}, which can significantly affect the SN environments. The specific adjustments needed in the feedback models however requires a more careful scrutiny of all physics involved in the simulation, and will be addressed in future work. Some aspects of the simulation-observation comparison may need to be better tailored to isolate differences due to feedback alone. E.g. the simulated dwarf galaxy system is about a factor of 100 lower in stellar mass than M33 \citep{Corbelli2003}, and although we only include the pre-starburst phase where the specific star-formation rate is similar to M33, the overall ISM phase distribution and porosity could still be quite different. Inclusion of the warm and hot ISM tracers in the observations as mentioned earlier can provide a fairer comparison of the low-density tail with simulations. Higher-resolution observations (e.g. Fig \ref{fig:zoomin}) will better clarify whether some of the WRs near high density peaks are located in low-density cavities even in projection.

The exercise here demonstrates a novel and insightful line of comparison between observations and simulations to validate stellar feedback models. The method of tracking locations of SNe (or their progenitors) and their ambient gas densities is easily done with multi-wavelength observations and hydrodynamical simulations of galaxies, as shown here. In upcoming papers, we will continue such comparisons with simulations with varying prescriptions of stellar feedback \citep[e.g photoionization, winds, SNe, runaways,][]{Rathjen2021, Andersson2020, Steinwandel2023} in order to better understand which implementation is most consistent with the SNe and ISM observations. We also encourage simulators to carry out their own comparisons using data behind the Figures \ref{fig:3paneldensities} and \ref{fig:3paneldensities_rsgs}, which will be available with the journal publication. 

\subsection{Comparison with SN and SNR studies} \label{sec:compareSNstudies}

While our paper was focused specifically on the ISM properties and implications for feedback, we note the parallels of our work with past studies of star-formation rates around SNe, which show that SE-SNe (Type Ib/c, Type IIb) are more likely to be near bright \hii\ regions than Type II or Type Ia SNe \citep[e.g.][]{Kelly2008, Anderson2008, Anderson2012, Galbany2016b, Galbany2016, Galbany2018, Kunc2018, Cronin2021}. The trend has been attributed to an increasing progenitor mass range going from Type II to SE progenitors, hence their association with younger star-forming populations. Similar results were also found with molecular gas, which correlates with recent star-formation rate, in kpc-scale maps of galaxies \citep{Galbany2017} as well as the GMC scale studies of MC22. Our work also shows that RSGs, which are expected to be progenitors of Type II SNe, are less closely associated with star-forming H$_2$ gas than WRs. Similar to our Figure \ref{fig:mocks}, Figures 10 and 11 in MC22 show that Type Ib/c’s can be reproduced by distributions drawn from the H$_2$ density peaks, while Type II SNe closely follow the stellar distribution traced by near-IR maps. A large fraction (53\%) of the MC22 sample also has CO non-detections similar to our findings in Figures \ref{fig:3paneldensities} and \ref{fig:3paneldensities_rsgs}. Zoom-in images of the SNe in Figures 8 in MC22 display the wide variety of environments apparent even at 150 pc, with SNe evolving in bright dense gas peaks, some in regions of lower density, clouds and general gas deficit, and some in outright voids in the ISM. 

We also note the parallels of our work with \cite{Kangas2017}, who compared the locations of massive stars (including RSGs and WRs) on H$\alpha$ images of LMC and M33, and compared the correlations with that observed for SNe vs H$\alpha$ observed in nearby galaxies. They find that WRs and SE-SNe have similar and stronger spatial correlation with H$\alpha$ emission (and thus recent star-formation) compared to RSGs and Type II SNe. 

Overall, it is encouraging that the results with massive star progenitors and SNe converge, validating the use of these massive stars as a local analog of SN sites, where we have substantial detail compared to distant galaxies, allowing us to  deconstruct their environmental details. We believe higher resolution studies of both our Local Group stars, and PHANGS SNe promise to provide even more subtle details involving high mass/short-lifetime progenitors.

A final comparison we make is with the statistics of known SNRs interacting with molecular gas. Many well-studied interacting SNRs exist in our Galaxy such as IC443, W44, W28 and W51C, visible in broadened spectral line and continuum emission across the EM spectrum \citep[e.g. see][and references therein]{Slane2015}. These provide the most direct evidence that SNe interact with molecular clouds. The exact fraction of interacting SNRs in our Galaxy however is highly uncertain due to the unknown classifications of most SNRs, and unknown completeness of the highly heterogeneous collection of Galactic SNRs. To our knowledge, the most systematic study of interacting SNRs was done by \cite{Kilpatrick2016}, who looked for broadened CO emission towards a subset of SNRs. They found 17/46 SNRs ($\sim$37\%), or 7/15 ($\sim$47\%) if restricted to SNRs with known CC classification, showing potential evidence of interaction. This is roughly consistent with our 39\% SNRs in a detected H$_2$ pixel, though it is much lower if we consider SNRs within H$_2$ pixels within a cooling radius (66-80\%). While this may suggest the interacting SNRs in our Galaxy may be an undercount, we are only going by positional coincidence. It is possible the SNRs may be in the foreground or background of the cloud. These SNRs would be interesting targets for follow-up with more sensitive high-resolution ALMA observations, where signatures of interaction could be inferred from moment data and their correlation with optical line/X-ray luminosities. We leave this for a future work.

\section{Conclusion} \label{sec:conc}
With ambient density being a decisive parameter for SN thermal and momentum feedback, we have embarked on a much-needed detailed observational campaign to map the gas distribution around SNe by employing a novel and unique strategy -- targeting evolved massive stars that are soon to explode near their present location, and their cold ISM environments. This not only yields an order-of-magnitude larger sample of `explosion sites' of stars than possible with a SN survey in the nearby universe, but also exploits the proximity of Local Group galaxies (in our case M33), which has not only cloud-scale molecular gas maps, but also atomic gas maps, giving estimates of the \hi, H$_2$ and total cold ISM densities in the 50 pc environments of massive stars. 

Specifically, we use published catalogs of evolved massive stars such as red supergiants (RSGs) and Wolf-Rayet (WRs) stars that have been connected to SN explosions. These stars will have lifetimes of roughly 0.1--1 Myr, and thus explode within a few pcs of their current location. While RSGs plausibly trace locations of future Type II SNe, WRs to first order trace where the most massive ($>$30 \Msun) and youngest stellar populations are located and may explode in the future as stripped-envelope SNe and gamma-ray bursts. We also include supernova remnants (SNRs), verified by high-quality multi-wavelength data, as a tracer of where stars have already exploded. 

To trace the cold ISM, we use new ALMA ACA maps of CO(2-1)-traced H$_2$ and 21 cm atomic \hi in the inner projected $\sim 3\times5$ kpc$^2$ of M33 (E. Koch et al., in prep), complete down to $\sim$1--2\Msunpc, and at 12\arcsec\ resolution ($\sim$50 pc). We then assess the relative cumulative distribution function of their densities, and assess the physical and statistical nature of these distributions using multi-wavelength star-formation maps. 

Our investigation revealed the following about where stars explode in the ISM:
\begin{enumerate}
\item We find a significant fraction of high-mass stars located in pixels with detectable H$_2$, and higher the mass, higher is the average ambient density. This is evidenced by 55\% of WRs exploding in pixels with detectable H$_2$, compared to 27-32\% for RSGs and SNRs (Figures \ref{fig:3paneldensities},\ref{fig:3paneldensities_rsgs}, \ref{fig:ntot_progmass}, Table \ref{tab:stats}), as well as from statistical differences in the shapes of the H$_2$ CDFs (Section \ref{sec:stats}, \ref{sec:sfrtracers}). This trend is also observed within RSGs: more luminous RSGs with log$(L/L_{\odot})>5$, corresponding to masses $>$15 \Msun, have a higher fraction of stars (44\%) exploding in denser gas than the bulk RSG population.
\item With our observations, we estimate that $\sim$45\% for WRs, and 56-77\% for the RSGs (depending on mass) are located in pixels with no detectable  H$_2$. This is similar to the result of \cite{MC22} who found that 50\% of Type II SNe explode outside GMCs at 150-pc resolution. Analysis of the star-molecular cloud separation distances in Section \ref{sec:reasonforlowdensities} indicate that pre-SN and prior-SN clearing of H$_2$ clouds is the likely cause for most of the non-detections of WRs. Both pre-SN/prior-SN clearing, and stellar motion away from birth sites could be responsible for the H$_2$ non-detections around the lower-mass RSGs.

\item For the first time, we have measured both molecular \emph{and} atomic gas densities around SN progenitors at 50 pc resolution (Figures \ref{fig:3paneldensities}+\ref{fig:3paneldensities_rsgs}). Unlike H$_2$, we find that majority of our stars ($\gtrsim$90\%) are located in pixels with detectable atomic gas. Of the H$_2$ non-detected population, 83-94\% of WRs, RSGs and SNRs are in pixels with detectable \hi, implying that these explosions will more likely occur in the lower density, atomic-gas dominated intercloud media.

\item We verified that the observed ISM density distributions around WRs and RSGs $>$15 \Msun are statistically different from random alignment of stars with respect to molecular clouds (Section \ref{sec:stats}, \ref{sec:sfrtracers}), and consistent with ISM densities where stars of similar ages would be expected to explode (Figures \ref{fig:mockmaps} and \ref{fig:mocks}). Ambient densities of the lower mass RSGs and SNRs are marginally different from the bulk densities, and appear to follow the near-infrared light in M33 tracing bulk stellar mass.

\item The H$_2$ detection statistics around stars is sensitive to the spatial resolution of the ISM maps, similar to the conclusion reached by \cite{MC22}. The H$_2$ non-detection fractions decrease from 45-77\% at 50 pc resolution to 0\% at $\sim$0.6 kpc resolution (Figure \ref{fig:aperture}). At our highest resolution H$_2$ data (at $\sim$1.8 pc) around a single WR star, we find a 20-pc scale cavity that was not visible in the 50 pc-scale ACA map (Figure \ref{fig:zoomin}) These reveal the importance of pre-SN/prior-SN feedback, possible evidence of triggered star-formation, and the need for high-resolution ISM maps for revealing such ISM substructure moving forward.

\item Our measured densities can be a useful observational anchor on the placement of SNe in ISM subgrid models in hydrodynamical simulations of galaxies, particularly to set limits on the fraction that is uncorrelated with density peaks. An example comparison is done with the high-resolution dwarf galaxy system by \cite{Lahen2020} in Figure \ref{fig:simvsobs}, showing some subtle differences between observed and simulated SN density distributions. Future observations at higher resolution, a wider stellar mass range of galaxies, and comparison with simulations more tailored to the observed galaxy type, can more specifically constrain the underlying stellar feedback models, and the necessary adjustments.

\end{enumerate}

\section*{Acknowledgements}
We acknowledge helpful feedback from the anonymous referees, and useful discussions with Drs. Thorsten Naab, Eve Ostriker and Todd Thompson on stellar
feedback models. LC grateful for support from the National Science Foundation in the form of grants AST-2107070 and AST-2205628. V. V. acknowledges support from the scholarship ANID-FULBRIGHT BIO 2016 - 56160020, funding from NRAO Student Observing Support (SOS) - SOSPADA-015, and partial support from NSF-AST1615960. The National Radio Astronomy Observatory is a facility of the National Science Foundation operated under cooperative agreement by Associated Universities, Inc. This paper makes use of the following ALMA data: 2017.1.00901.S; 2019.1.01182.S. ALMA is a partnership of ESO (representing its member states), NSF (USA) and NINS (Japan), together with NRC (Canada), MOST and ASIAA (Taiwan), and KASI (Republic of Korea), in cooperation with the Republic of Chile. The Joint ALMA Observatory is operated by ESO, AUI/NRAO and NAOJ 
Support for K.F.N was provided by NASA through the NASA Hubble Fellowship grant (HST-HF2-51516) awarded by the Space Telescope Science Institute, which is operated by the Association of Universities for Research in Astronomy, Inc., for NASA, under contract NAS5-26555.
NL acknowledges the computational resources granted by CSC -- IT Center for Science Ltd. in Finland and The Max Planck Computing and Data Facility (MPCDF) in Garching, Germany. SS and NMP acknowledge support provided by the University of Wisconsin - Madison
Office of the Vice Chancellor for Research and Graduate Education with funding from the Wisconsin Alumni Research Foundation, and the NSF Award AST-2205630. This research made use of Photutils, an Astropy package for
detection and photometry of astronomical sources (Bradley et al.\ 2023).  The research made use of the z0MGS GALEX+WISE data \citep{zomgsdoi}, which is a GALEX+WISE matched resolution image atlas for around 10,000 nearby galaxies.This dataset or service is made available by the Infrared Science Archive (IRSA) at IPAC, which is operated by the California Institute of Technology under contract with the National Aeronautics and Space Administration.

%

\vspace{5mm}
\facilities{VLA, ALMA, Mayall, UKIRT}


\software{\textbf{\texttt{CASA} \citep{CASA}, \texttt{Astropy} \citep{astropy:2013, astropy:2018, astropy:2022}}, \texttt{matplotlib} \citep{matplotlib}, \texttt{scipy} \citep{scipy}, \texttt{numpy} \citep{numpy}, \texttt{photutils} \citep{photutils}.}



\appendix
\section{Calculating the spatially-varying volume density} \label{appendix:volumedensity}
\begin{figure}
    \centering
    \includegraphics[width=\columnwidth]{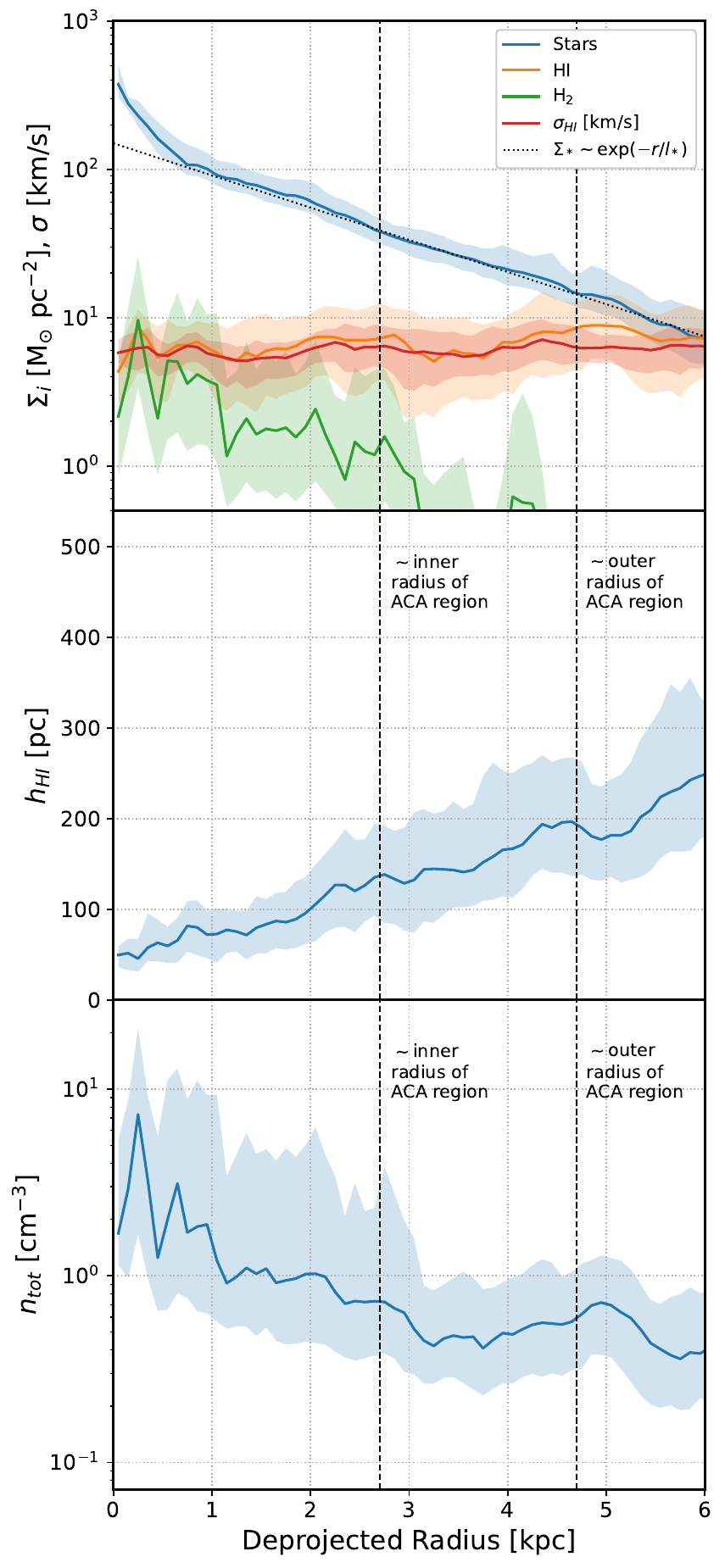}
    \caption{Variation in the \hi, H$_2$, and stellar mass surface densities and \hi velocity dispersion (\emph{top}), scale height (\emph{middle}) and total \hi + H$_2$ volume density (\emph{bottom}) versus deprojected radius of M33. Dashed lines represent the approximate extents of the ACA region in Figure \ref{fig:maps} in radius units. The dotted line in the top panel represents an exponential disk model, assumed in the text. Shaded regions represent the 16th-84th percentile range, while solid lines represent the median. See Appendix \ref{appendix:volumedensity} for details. }
    \label{fig:radialprofiles}
\end{figure}

Given measurements of $\Sigma_{HI}$ and $\Sigma_{H_2}$, the average total (\hi + H$_2$) density ($n_{\mathrm{tot}}$, in units of atoms per cm$^{-3}$) can be obtained as 
\begin{equation}
n_{\mathrm{tot}} = \frac{1}{2\mu m_p}\left(\frac{\Sigma_{HI}}{h_{HI}} + 2\frac{\Sigma_{H_2}}{h_{H_2}}\right),
\label{eq:ntot}
\end{equation}

where the factor of 2 with $\Sigma_{H_2}$ accounts for 2 atoms of H per molecule, $\mu=1.4$, and $m_p = 1.6 \times 10^{-24}$ g. The underlying assumption is that both gas phases are uniformly distributed along each column, characterized by scale heights $h_{H_I}$ and $h_{H_2}$. While obviously the real feedback-driven ISM is inhomogeneous and turbulent, this simple parameterization of the line-of-sight gas gives an easy-to-interpret average density per pixel as a function of the observed projected surface density in order to enable first-order comparisons with simulations.

The scale height of \hi ($h_{HI}$) can be estimated assuming the disk of M33 is in a state of dynamical equilibrium, i.e., the gravitational weight of the inner disk due to stellar and gaseous components is balanced by thermal and turbulent pressure of the diffuse gas driven by star-formation and stellar feedback \citep{OML10, OS11, SO12}. Numerical simulations have verified that within an orbital time, star-forming disks reach such quasi-steady equilibrium properties due to self-regulation of star formation by feedback \citep[e.g.,][]{Kim2011, Kim2013, Ostriker2022}, and the resulting star-formation rates and related observables are broadly consistent with observations \citep[e.g.,][]{Gallagher2018, Sun2020b, Sun2023}. This formalism therefore provides a unique way to obtain the \emph{spatially-varying} scale height and line-of-sight volume density of the atomic ISM for a nearly face-on galaxy like M33, as a function of locally measured quantities such as the gas and stellar mass surface density, kinematics, and star-formation rates. 

For our purpose, we closely follow the arguments in \cite{Utomo2019} to obtain the scale height, with a few minor modifications. We assume that the vertical pressure in the \hi gas in M33 is related to its velocity dispersion and density by
\begin{equation} \label{eq:PHI}
P_{HI} = \rho_{HI}\ \sigma_{HI}^2.
\end{equation}
In a disk under dynamical equilibrium, the midplane pressure is given by,
\begin{equation} \label{eq:Ptot}
P_{\mathrm{tot}} = f_d\frac{\pi G \Sigma^2}{4}\left[(2-f_d) + \Biggl\{(2-f_d)^2 + \frac{32\sigma^2\rho_{\mathrm{sd}}}{\pi^2 G \Sigma^2}\Biggr\}^{1/2}\right]
\end{equation}
in the form given in Eq (35) of \cite{Kim2011}. Here $\Sigma = \Sigma_{HI} + \Sigma_{H_2}$ is the total gas surface density, and $f_d$ is the diffuse gas fraction, i.e., $f_d = \Sigma_{HI}/\Sigma$. The gas velocity dispersion $\sigma$ is assumed to be equal to $\sigma_{HI}$. The total stellar + dark-matter density is $\rho_{sd}$, but since we are only interested in the inner few kpc of M33, we can approximate $\rho_{sd} \approx \rho_{*}$, i.e., only the stellar density \citep{Corbelli2003}. 

The stellar volume density, $\rho_{*}$, can be computed from a near-IR map of M33. Similar to \hi\ and H$_2$, we assume $\rho_{*} \approx \Sigma_{*}/2h_*$, where $h_{*}$ is the stellar scale height obtained by assuming a flattening ratio between the stellar scale lengths and heights, $l_*/h_* = 7.3$ \citep{Kregel2002}, where we assume the scale length $l_* \approx 2$ kpc based on photometric observations in M33 \citep{Barker2007}, and also in Figure \ref{fig:radialprofiles}. This gives $h_* = 0.27$ kpc. We compute $\Sigma_{*}$ from the \emph{WISE} W1 (3.4 $\mu$m) map following the conversion given in \cite{Leroy2019}
\begin{equation}
\frac{\Sigma_{*}}{\mathrm{M}_{\odot} \mathrm{pc}^{-2}} \approx 
330 \left(\frac{M/L}{0.5}\right)\left(\frac{I_{3.4\mu m}}{\mathrm{MJy}\ \mathrm{sr}^{-1}}\right), 
\end{equation}
where $M/L$ is the mass-to-light ratio in the 3.4 $\mu$m range. The $M/L$ is expected to correlate with the underlying age of the stellar population, and thus vary across the galaxy \citep{Stanway2018}. We account for this based on the empirical relation with the specific star-formation rate given in Eq (24) in \cite{Leroy2019} as
\begin{equation}
M/L =     \begin{cases}
    0.5 & \text{if}\ Q<a \\
    0.5 + b(Q-a) & \text{if}\ a<Q<c \\
    0.2 & \text{if}\ Q>c
    \end{cases}
    \label{eq:m2l}
\end{equation}
Here $Q = \mathrm{log_{10}} (L_{\mathrm{FUV}}/L_{W1})$ is the ratio between the FUV and W1 luminosities, roughy equivalent to the specific star-formation rate, $a = -2.5$, $b = -0.167$ and $c=-0.7$. Both FUV and W1 maps were convolved to the resolution and pixel scale of the ACA map before the conversions were applied. Within the ACA region, the $M/L$ is mostly 0.2 \Msun/L$_{\odot}$, consistent with observations of galaxies with similar specific star-formation rates as M33 \citep{Leroy2019}.

With the above quantities in place and assuming $\rho_{HI} = \Sigma_{HI}/2 h_{HI}$, we obtain the scale height $h_{HI}$ by equating Eq \eqref{eq:PHI} and \eqref{eq:Ptot}, giving
\begin{equation}
    h_{HI} = \frac{1}{2}\frac{\Sigma_{HI}\sigma_{HI}^2}{P_{\mathrm{tot}}}
\end{equation}
providing a spatially-varying \hi scale height as a function of the local measured atomic, molecular and stellar mass surface densities, and the atomic line-width maps. We have verified from simulations of \cite{Ostriker2022} that these scale height equation provide a reasonable estimate (within 10-20\%) of the average density at 50 pc resolution. Further accuracy can be obtained in future work by including the mass of hot gas from X-ray observations.

The molecular scale height, $h_{H_2}$, however is less trivially obtained as the gas is primarily locked in gravitationally-bound clouds that are not necessarily in pressure equilibrium with the disk, and equations of hydrostatic balance (similar to Eq \eqref{eq:Ptot}) have been shown to over-estimate the actual scale height of the clumpy molecular disk \citep{Jeffreson2022}. For simplicity, we assume $h_{H_2}=50$ pc, which is consistent with measurements made in the Milky Way \citep[e.g.][]{Marasco2017}. This may be slightly smaller than the expected value for M33, which has a much smaller stellar mass (and thus smaller gravitational potential) than the Milky Way. We found that $n_{\mathrm{tot}}$ obtained by  assuming $h_{H_2}=50$ pc is on average only about 40\% larger than the limiting case of $h_{H_2}=h_{HI}$. This is much smaller than the orders of magnitude variation in densities expected across the disk in the ACA region, so we believe $h_{H_2}=50$ pc is a reasonable value for M33.
\begin{figure*}[t]
\gridline{\fig{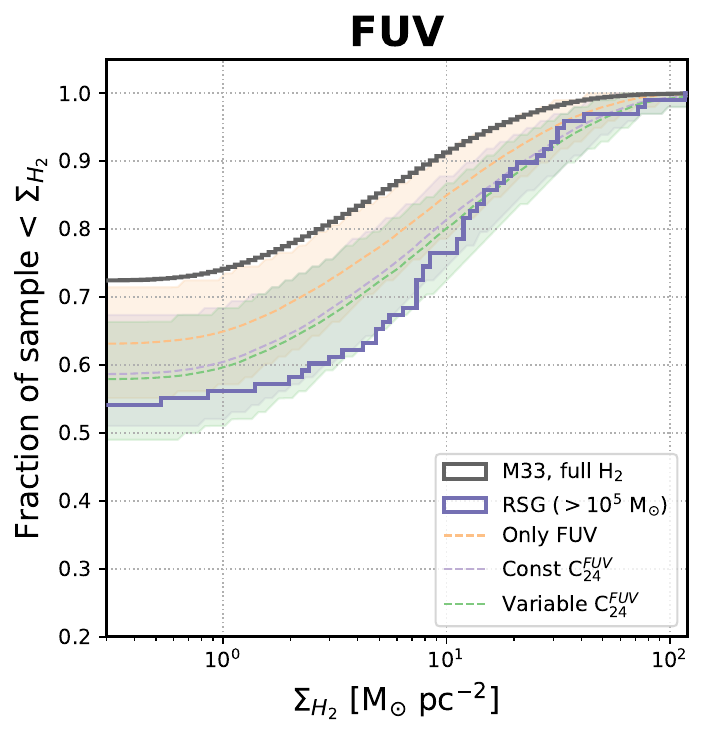}{0.5\textwidth}{(a)}
          \fig{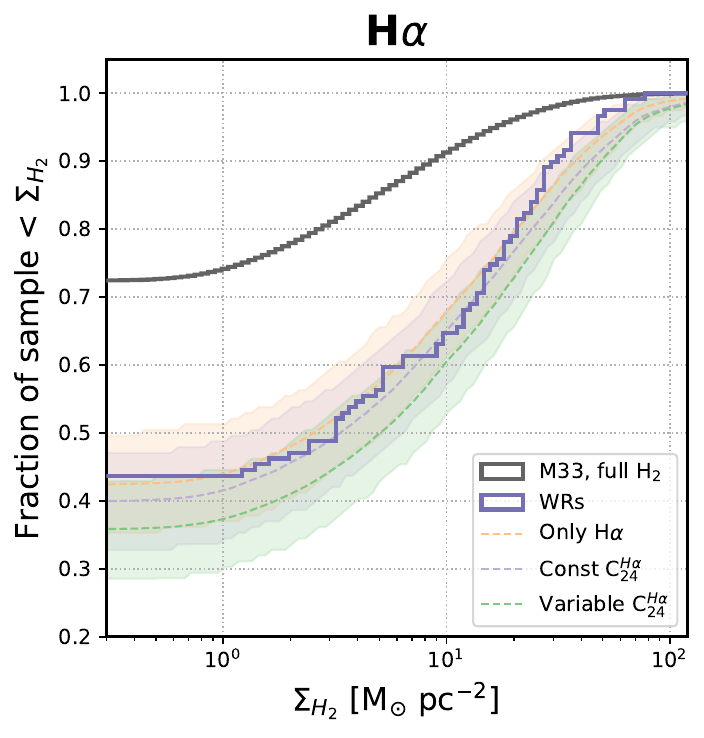}{0.5\textwidth}{(b)}}
    \caption{The H$_2$ surface density CDFs similar to Figure \ref{fig:mocks} (dashed line represents median, shaded region represents 95\% interval), except showing the different 24 $\mu$m extinction-corrections for FUV (\emph{left}) and H$\alpha$ (\emph{right}) as discussed in Section \ref{sec:extcorrfuvha}. ``Only FUV/H$\alpha$'' refers to the maps without extinction correction, ``Const'' C$^{FUV,H\alpha}_{24}$ refers to a constant value of the extinction-correction coefficient, and ``Variable'' C$^{FUV,H\alpha}_{24}$ refers to the coefficient scaled by the WISE W1-band maps according to Belfiore et al (2023) to account for dust-heating by old stellar populations. For simplicity, since the focus is on differences between the prescriptions, we only show the high-mass RSGs on the FUV panel and WRs on the H$\alpha$ panel, since these were most consistent with the respective wavelengths in Figure \ref{fig:mocks}.}
    \label{fig:fuvha}
\end{figure*}
Figures \ref{fig:radialprofiles} shows the variation in the relevant quantities in the above equations as a function of deprojected radius of M33 (note that the calculations in the paper are pixel-by-pixel, not radial; we just show these quantities vs radius to give a succinct idea of their variation). Deprojection is done assuming the center of M33 is at RA(2000) = 01h:33m:50.89657s, Dec(2000)= +30$^{\circ}$:39$^{\prime}$:36.630403$^{\prime\prime}$, an inclination angle $55^{\circ}$, and position angle of $200^{\circ}$ \citep{Koch2018}. The profiles are plotted in radial bins of $\Delta r = 100$ pc. Within each bin, for the stellar and gas mass densities, we estimate a median, 16th, and 84th percentile density, multiply by the area covered by the \emph{detected} pixels to get the total mass, and divide this mass by the total area inside the $\Delta r$ to get the radial surface density. For the volume density and scale height, we show their median and 16th-84th percentile range of their pixel values within each $\Delta r$.

In particular, we highlight the profiles of the scale height and total volume density in Figure \ref{fig:radialprofiles}. The \hi\ scale height as shown in the middle panel increases with distance from the center because $\Sigma_*$, which provides the dominant contribution to the gravitational potential of the disk within 5 kpc, drops with radius as $\sim$$\mathrm{e}^{-r/l_*}$, while the diffuse gas properties that provide outward pressure (i.e., $\Sigma_{HI}$, $\sigma_{HI}$) remain relatively flat. The scatter in the scale height is primarily due to region-to-region variation in $\Sigma_{HI}$, $\Sigma_{H_2}$, $\Sigma_*$, and $\sigma_{HI}$, which we are able to capture with this above approach (as opposed to assuming a constant value throughout the disk). The final panel in Figure \ref{fig:radialprofiles} shows the total resulting volume density (assuming h$_{H_2}$=50 pc) vs deprojected radius from Eq \eqref{eq:ntot}. The scatter in density folds in the scatter from the above quantities, but overall the density decreases with distance from the center. The density within 1 kpc of the center has particularly large scatter and is higher than at outer radii due to more significant contribution from molecular gas (whose surface density also drops off exponentially with radius as seen in the top panel).  Both the scale height and the density values are roughly consistent with values in the Milky Way \citep{Dickey1990} and other edge-on galaxies \citep[e.g.,][]{Patra2019, Randria2021, Zheng2022}, as well as simulations of the multiphase ISM in realistic galaxy environments \citep[e.g.,][]{Hill2012, Hopkins2012, Kim2017b}.

\section{Extinction-corrected FUV and H$\alpha$ maps} \label{sec:extcorrfuvha}
We extinction-corrected the FUV and H$\alpha$ maps using MIPS 24 $\mu$m, as prescribed in \cite{Belfiore2023} (hereafter B23). We first convolve and regrid all maps to 12\arcsec 
 resolution and 1\arcsec\ pixel scale as the ACA maps, and convert them to units of luminosity. We then convert them to units of hybrid star-formation rate as given in Eq (4) as
\begin{equation}
SFR_{\mathrm{FUV+24}} = C_{\mathrm{FUV}}L_{\mathrm{FUV}} + C^{\mathrm{FUV}}_{24} L_{24}
\end{equation}
where SFR$_{FUV+24}$ is the 24$\mu$m corrected FUV-based star-formation rate in units of \Msun yr$^{-1}$, L$_{FUV}$ is the FUV luminosity in erg s$^{-1}$ and L$_{24}$ is the 24$\mu$m MIPS luminosity in erg s$^{-1}$, $C_{\mathrm{FUV}}=10^{-43.42}$ \cmipsunit given in Table 2 of B23 and derived in \cite{Leroy2019}, and $C^{\mathrm{FUV}}_{24}$ is the 24$\mu$m correction factor for FUV. Similarly for H$\alpha$ we follow Eq (5) in B23 as 
\begin{equation}
SFR_{\mathrm{H\alpha+24}} = C_{\mathrm{H\alpha}}L_{\mathrm{H\alpha}} + C^{\mathrm{H\alpha}}_{24} L_{24}
\end{equation}
where the terms carry the same meaning as above, except for H$\alpha$ maps. We use $C_{\mathrm{H\alpha}}=10^{-41.26}$ \cmipsunit as given in Table 2 of B23, derived in \cite{Calzetti2007}.

The correction factors C$^{FUV}_{24}$ and C$^{\mathrm{H\alpha}}_{24}$ were adjusted by B23 to more accurately account for dust heating by lower-mass stellar populations. We use the calibration form of Eq 7 in B23
\begin{equation}
\mathrm{log}\ C^{i}_{24} = \begin{cases}
    \mathrm{log}\ C_{\mathrm{max}} \\ \quad +\ a_1 (\mathrm{log}\ Q - \mathrm{log}\ Q_{\mathrm{max}}) & \text{if}\ Q<Q_{\mathrm{max}} \\
    \mathrm{log}\ C_{\mathrm{max}} & \text{if}\ Q>Q_{\mathrm{max}} \\
    \end{cases}
    \label{eq:logci24}
\end{equation}
where the subscript $i$ refers to H$\alpha$ or FUV, and $Q$ refers to the quantities log$(L_{FUV}/L_{W1})$ and log$(L_{H\alpha}/L_{W1})$, as already defined in Eq \ref{eq:m2l}. Based on Table 3 in B23, for H$\alpha$, we adopt $a_1=0.45$, log $C_{\mathrm{max}}=-42.88$, and  log $Q_{\mathrm{max}}=-1.52$. Similarly, for FUV, we adopt $a_1=0.23$, log $C_{\mathrm{max}}=-42.73$, and log\ $Q_{\mathrm{max}}=0.6$. We note that B23 estimated these scaling relations for log $C^{i}_{24}$ in the WISE W4 (22 $\mu$m) band, which was different from measurements in MIPS 24 $\mu$m by 0.08 dex. As a result, we add 0.08 to log $C^{i}_{24}$ in Eq \ref{eq:logci24} since we are working with MIPS 24$\mu$m maps. 
\begin{figure}
    \includegraphics[width=\columnwidth]{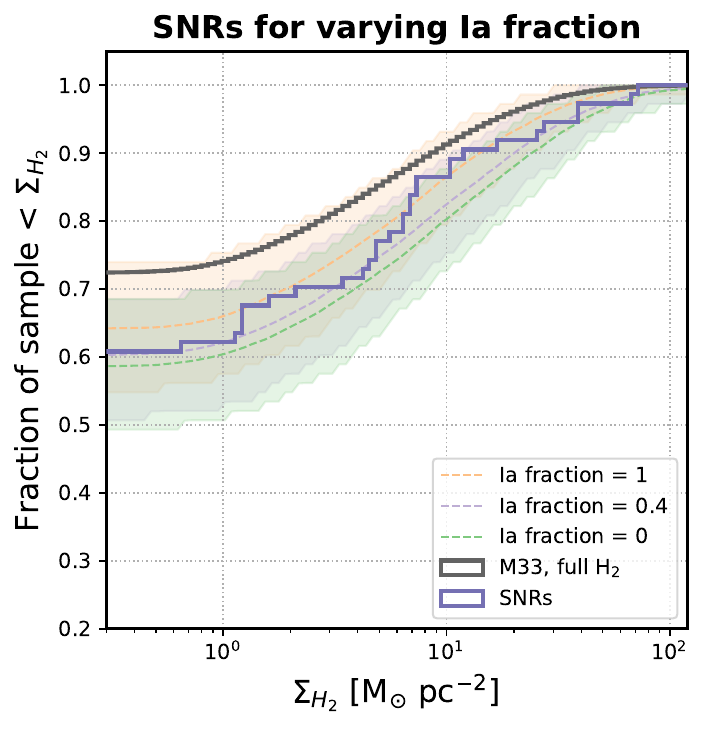}
    \caption{The H$_2$ surface density CDFs for mock population of stars drawn from near-IR and FUV maps meant to replicate SNRs with some combination of core-collapse and Type Ia SNe. This combination is represented by the `Ia fraction', the fraction of the population drawn from near-IR (rest drawn from FUV). See Section \ref{sec:snria} for details.}
    \label{fig:snrsia}
\end{figure}

Figure \ref{fig:fuvha} shows that CDFs exhibit small variations, mostly within error-bars of each other, for the different prescriptions discussed above. Including the 24$\mu$m correction, the median fraction of the H$\alpha$-drawn star sample occurring in detectable H$_2$ is higher by 8\% and the fraction of FUV-drawn stars occurring in H$_2$ is higher by 7\%, compared to the maps without 24 $\mu$m correction. The higher fraction is due to the infra-red capturing more obscured regions in M33, which are expected to be associated with denser gas and younger, more obscured stellar populations. Interestingly, the WR population in Figure \ref{fig:fuvha} is more consistent with the H$\alpha$-population without extinction correction. This indicates that the spatial distribution of WRs is similar to the unobscured, ionizing OB stars, and reinforces that we are not missing a significant fraction ($\lesssim$7\%) of WRs in highly obscured gas clouds. For completeness, we also show the CDFs for FUV and H$\alpha$ maps with a constant correction factor, as has been traditionally used in the literature. For this purpose, we used the constant values given in Table 2 of B23, specifically C$^{H\alpha}_{24}=10^{-42.78}$ \citep{Calzetti2007} and C$^{FUV}_{24}=10^{-42.83}$ \citep{Hao2011, Leroy2019}. The difference in the corresponding CDFs lies in between the no-extinction and variable coefficient cases as seen in Figure \ref{fig:fuvha}.

As explained in Section \ref{sec:sfrtracers}, these differences are not consequential to the main science result, that the densities of the observed RSG and WR population are not randomly drawn, and are roughly consistent with their progenitor population.

\section{The contribution of Type Ia SNe in the SNR densities} \label{sec:snria}
We assess how the observed CDF of H$_2$ densities of SNRs in the last row of Figure \ref{fig:mocks} compares with model CDFs of H$_2$ densities drawn from a combination of SF-tracer maps that represents some proportion of SNe Ia (with low-mass progenitors) and core-collapse SNe (with high mass progenitors) in the sample. We know from \cite{Koplitz2023} that the Type Ia fraction is potentially 40\% of the SNRs in the ACA region based on star-formation histories. We repeat the experiment in Section \ref{sec:sfrtracers} and Figure \ref{fig:mocks} for SNRs, except now assuming that some fraction of each randomly-drawn sample of SNRs comes from the near-IR map (representing old progenitors of Type Ia SNe) and the rest from the FUV map (representing the massive progenitors of core-collapse SNe). 
\begin{figure}
\centering
\includegraphics[width=\columnwidth]{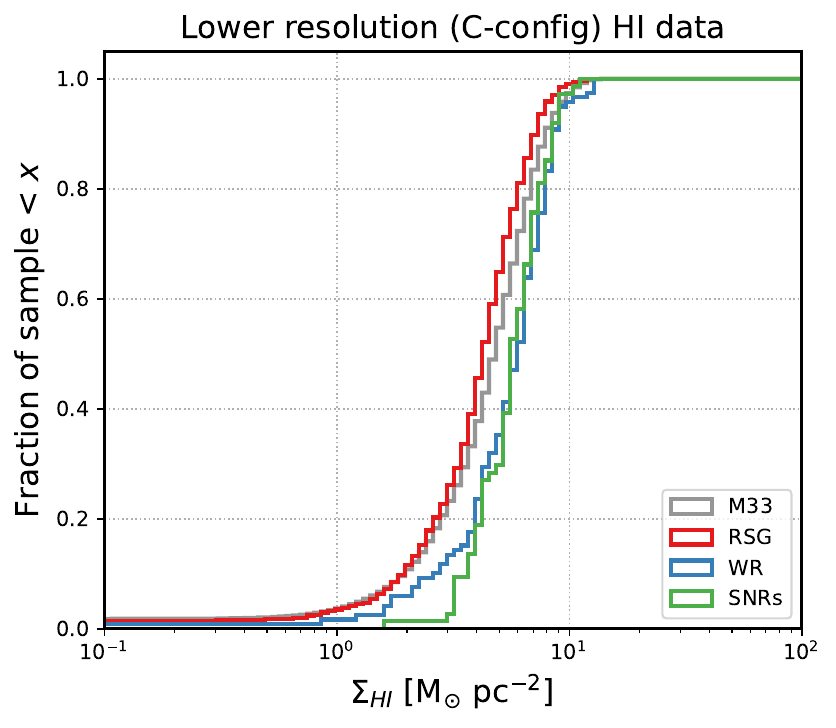}
    \caption{Same as Figure \ref{fig:3paneldensities}, but using the lower resolution C-configuration \hi data.}
    \label{fig:Cconfig}
\end{figure}

The results are shown in Figure \ref{fig:snrsia}. While all three models are consistent with each other within the shaded error regions, we do find the median model CDF of H$_2$ densities, where we draw 40\% of the sample from the near-IR to represent Type Ia SNe, is slightly more consistent with the observed CDF (purple) than the two extreme scenarios, i.e. all SNRs are Type Ia (i.e. drawn from near-IR) or all are core-collapse (drawn from FUV). This points to some Type Ia contribution to the observed H$_2$ density CDF of SNRs. Revisiting this experiment with the full M33 SNR population (which we can utilize once the full CO map of M33 is available) will help reduce the shaded error region, and put stronger constraints on the Ia contribution to the observed H$_2$ densities.

\section{Lower resolution HI data} \label{sec:Cconfig}
Figure \ref{fig:Cconfig} shows the same result as the middle panel of Figure \ref{fig:3paneldensities}, but with the lower-resolution C-configuration ($\sim$21\arcsec $\sim$86 pc) \hi data from \cite{Koch2018}. The increased surface brightness sensitivity of the lower resolution data leads to a lower fraction of non-detections in the 3 categories of stars, specifically 2\%, 3\% and 0\% for WRs, RSGs and SNRs respectively. This provides a minor, but an important point to note for atomic gas densities measured by interferometry. The higher-resolution B+C+D configuration VLA data has higher RMS noise than the C+D data, so the fraction of non-detections around stars is higher in B+C+D than C+D. Thus, the 10\% of stars in Fig \ref{fig:3paneldensities} that we claim are \hi non-detections are not completely devoid of any \hi, but rather may still have lower density \hi along the line of sight below the BCD data sensitivity.

\bibliography{resubmit_main_revised}{}
\bibliographystyle{aasjournal}

\suppressAffiliationsfalse
\allauthors


\end{document}